\documentclass{aa}  

\usepackage{lastpage}
\usepackage{graphicx}
\usepackage{txfonts}
\usepackage{natbib}
\usepackage[colorlinks=true,
    linkcolor=blue,
    filecolor=magenta,      
    urlcolor=blue,
    citecolor=blue]{hyperref}

\usepackage{xcolor}
\usepackage{scrextend}
\usepackage{siunitx}

\usepackage{bm}
\usepackage{amsmath,amssymb}
\usepackage{booktabs}
\usepackage{graphicx}
\usepackage{float}
\usepackage{comment}
\usepackage{diagbox}
\usepackage{tabularx}
\usepackage{threeparttable}
\usepackage{xspace}
\usepackage{placeins}
\usepackage{isotope}
\usepackage{layouts}
\usepackage{mathtools}

\def\ltsima{$\; \buildrel < \over \sim \;$}
\def\simlt{\lower.5ex\hbox{\ltsima}}
\def\gtsima{$\; \buildrel > \over \sim \;$}
\def\simgt{\lower.5ex\hbox{\gtsima}}

\usepackage{color}

\newcommand{\orcid}[2][0000-0000-0000-0000]{\href{https://orcid.org/#1}{#2}}

\newcommand{\Teff}{\ifmmode {T_{\rm eff}}\else${T_{\rm eff}}$\fi}
\newcommand{\Msun}{\xspace\ensuremath{M_\odot}\xspace}

\newcommand{\Rsun}{\ensuremath{R_\odot}}

\newcommand{\Zsun}{\ensuremath{Z_\odot}\xspace}
\newcommand{\Mdonor}{\ensuremath{M_\mathrm{donor}}\xspace}
\newcommand{\MBH}{\ensuremath{M_\mathrm{BH}}\xspace}
\newcommand{\Mmax}{\ensuremath{M_\mathrm{max}}\xspace}
\newcommand{\Mmin}{\ensuremath{M_\mathrm{min}}\xspace}
\newcommand{\posydon}{\texttt{POSYDON}\xspace}

\newcommand{\mesa}{\texttt{MESA}\xspace}
\newcommand{\chieff}{\ensuremath{\chi_\mathrm{eff}}\xspace}
\newcommand{\HMSHMS}{\texttt{HMS-HMS}\xspace}
\newcommand{\COHMSRLO}{\texttt{CO-HMS\_RLO}\xspace}

\begin{document}
   \title{A case for Case A: detailed look at binary black hole formation through stable mass transfer}
   \subtitle{}

    \author{\orcid[0000-0002-6842-3021]{Max M. Briel} \inst{1,2}\fnmsep\thanks{E-mail: max.briel@gmail.com}
            \and
            \orcid[0000-0003-1474-1523]{Tassos Fragos} \inst{1,2}
            \and
            \orcid[0000-0003-0648-2402]{Monica Gallegos-Garcia}\inst{3,4,5,6}
            \and
            \orcid[]{Anarya Ray}
            \inst{4,7}
            \and
            \orcid[0000-0002-0147-0835]{Michael Zevin}
            \inst{4,7,8}
            \and
            \orcid[0000-0002-6064-388X]{Abhishek Chattaraj}
            \inst{9}
            \and
            \orcid[0000-0001-5261-3923]{Jeff J. Andrews}
            \inst{9,10}
            \and
            \orcid[0000-0001-9236-5469]{Vicky Kalogera}
            \inst{3,4,7}
            \and
            \orcid[0000-0001-6692-6410]{Seth Gossage}
            \inst{4, 7}
            \and
            \orcid[0000-0003-1749-6295]{Philipp M. Srivastava}
            \inst{4,7,11}
            \and
            \orcid[0000-0003-0420-2067]{Elizabeth Teng}
            \inst{3,4,7}
    }
    \institute{Département d’Astronomie, Université de Genève, Chemin Pegasi 51, CH-1290 Versoix, Switzerland 
   \and
   Gravitational Wave Science Center (GWSC), Université de Genève, CH-1211 Geneva, Switzerland 
   \and
   Department of Physics and Astronomy, Northwestern University, 2145 Sheridan Road, Evanston, IL 60208, USA 
   \and
   Center for Interdisciplinary Exploration and Research in Astrophysics (CIERA), Northwestern University, 1800 Sherman Ave, Evanston, IL 60201, USA 
   \and
   Center for Astrophysics \textbar{} Harvard \& Smithsonian, 60 Garden St. Cambridge, MA, 02138, USA 
   \and
   Harvard Society of Fellows, 78 Mount Auburn Street, Cambridge, MA 02138, USA 
   \and
   The NSF-Simons AI Institute for the Sky (NSF-Simons SkAI), 172 E. Chestnut Street, Chicago, IL 60611, USA 
   \and
   The Adler Planetarium, 1300 South DuSable Lake Shore Drive, Chicago, 60605, IL, USA 
   \and
   Department of Physics, University of Florida, 2001 Museum Rd, Gainesville, FL 32611, USA 
   \and
   Institute for Fundamental Theory, 2001 Museum Rd, Gainesville, FL 32611, USA 
   \and
   Electrical and Computer Engineering, Northwestern University, 2145 Sheridan Road, Evanston, IL 60208, USA 
}

   \date{Received February, 2026; accepted XX, XX}
 
  \abstract
   {In isolated binary evolution, binary black hole (BBH) mergers are generally formed through stable mass transfer (SMT) or common envelope evolution. In recent years, the SMT channel has received significant attention due to detailed binary models showing increased mass transfer stability compared to previous studies.}
   {In this work, we perform a full zero-age-main-sequence to compact object merger analysis using detailed binary models at eight metallicities between $10^{-4}\,\Zsun$ and $2\,\Zsun$ to self-consistently model the population properties of BBH mergers in the SMT channel, determined their progenitor initial conditional, and investigate the binary physics governing their formation and metallicity dependence.}
   {We use the population synthesis code \posydon that incorporates detailed single-star and binary model grids to determine the population of BBH mergers from SMT. Using its extended grids of \mesa binary models, we determine the essential physics in the formation of BBH mergers.}
   {SMT produces BBH mergers predominantly from systems with $P_\mathrm{ZAMS} \leq 10$ days. In these systems, both the initial mass transfer between two stars and the subsequent interaction between the remaining star and the first-born BH take place while the respective donor star is on the main-sequence (Case A). We find a limited contribution from wider Case B/C systems. Without a natal kick, the SMT channel does not produce BBH mergers above $Z>0.2\,\Zsun$ due to orbital widening from stellar wind mass loss. The primary BH mass distribution shows a strong dependence on metallicity, while the mass ratio prefers unity independent of metallicity due to mass ratio reversal. Additionally, the \chieff distributions contain peaks at $\chieff=0$ and ${\sim}0.15$ of which the former disappears at high metallicities. A mass-scaled natal kick leave this sub-population unchanged but introduce a low-mass, unequal mass ratio sub-population that merges within the Hubble time due to their eccentricity.
   }
   {}


   \maketitle

\section{Introduction}

Over the past decade, the LIGO/Virgo/KAGRA (LVK) collaboration has published more than two hundred gravitational wave detections, offering a new window into the evolution of massive stars and binary evolution \citep{Abbott+19, Abbott+21, Abbott+23b, Abac+25}.
The majority of these events are binary black hole (BBH) mergers, and, due to the increasing sample size, population-level features are emerging, including overdensities in the primary black hole (BH) mass distribution near ${\sim}10\,\Msun$ and ${\sim}35\,\Msun$ \citep{Abbott+23a, Abac+25a}.
With GWTC-4 \citep{Abac+25b} more than doubling the BBH merger sample, these features have become more statistically significant than ever before. Interpreting the origin of the observational features to astrophysical processes has proven more complex than expected. Multiple formation channels, including isolated binary evolution, dynamical mergers in dense stellar environments, hierarchical mergers, chemically homogeneous evolution, and higher-order multiples, can contribute to the observed BBH merger population \citep[See][and references therein]{Mapelli+21a, Mandel+22}.
Each channel is expected to leave distinct imprints on the BBH merger population, but their relative contributions to the observed rates and population features remain uncertain \citep[e.g.][]{Wong+21, Zevin+21}.

BBH mergers from isolated binary evolution are expected to be shaped by interactions during the lifetime of the system \citep[e.g.][]{Mapelli+21a}. Binary interactions are essential for tightening the orbit, as two zero-age main-sequence (ZAMS) stars cannot fit within the separation required for a BBH merger to occur within the Hubble time ($t_\mathrm{Hubble} \sim 13.8$ Gyr). 
The common envelope \citep[CE; e.g.,][]{Belczynski+16a} and stable mass transfer \citep[SMT; e.g.,][]{vandenHeuvel+17} channels are generally identified as the dominant formation scenarios of merging BBHs within isolated binary evolution \citep[e.g.][]{Mapelli+20b}. In both, the binary typically first undergoes mass transfer with a stellar companion (the STAR+STAR phase). After the formation of the first BH (the STAR+BH phase), a binary in the CE channel undergoes dynamically unstable mass transfer and experiences a phase of CE evolution. If the envelope is successfully ejected, the resulting system typically has a period of less than a few days \citep{Ivanova+13, Ivanova+20}.
In the SMT channel, instabilities are avoided during the mass transfer, and the star reaches core-collapse without undergoing a CE. It generally leads to a variety of final orbital configurations, with the period only decreasing when the donor remains more massive than the accretor, though the exact boundary depends on the angular momentum loss during the mass transfer \citep[see][]{Shore+94, Renzo+19, vanSon+20, Willcox+23}.

Detailed modeling of the STAR+BH phase has shown that interactions are more stable than previously thought and can produce tight orbital configurations, even at solar metallicity \citep[e.g.][]{vandenHeuvel+17}.
Using a grid of detailed binary models, \citet{Marchant+21} showed that SMT leads to BBH mergers for a wide range of initial period and mass ratios at $0.1\Zsun$ for $\Mdonor=30\,\Msun$. This work demonstrated that SMT in the STAR+BH phase leads to BBH mergers in a narrow region along the boundary of unstable mass transfer. In this region (SMT-to-BBH-merger region), SMT shrinks the orbit efficiently due to the donor star mass (\Mdonor) being larger than the BH mass (\MBH) for a long duration during the mass transfer. This allows for sufficient orbital shrinkage before the formation of the second BH and the formed BBH system to merge within the Hubble time. \citet{Gallegos-Garcia+21} expanded the explored parameter space with additional masses and metallicities, and used these grids to model an astrophysical population. They found that SMT contributes significantly to the formation of BBH mergers, and affects both the rate and delay times of these systems, emphasizing the importance of understanding SMT in the context of gravitational wave sources.
Similar simulations by \citet{Klencki+26} showed that the region of SMT leading to BBH mergers is present over an even large range of star and BH masses, and metallicities. Through works such as these, the study of SMT during the STAR+BH phase has become increasingly more thorough, showing that SMT can robustly lead to BBH mergers within the Hubble time.

The application of the results for the STAR+BH phase within an astrophysical population of BBH mergers has been limited in the literature. One such example is the work by \citet{Gallegos-Garcia+21}, who used the outcome from the rapid population synthesis code \texttt{COSMIC} \citep{Breivik+20} to approximate the STAR+STAR phase before using detailed binary models for the STAR+BH phase, though at a limited metallicity and mass range. Building upon this, \citet{Briel+23} modeled the cosmic population of BBH mergers using the detailed \texttt{BPASS} binary models \citep{Eldridge+17, Stanway+18a} across a wide metallicity range, showing that SMT plays a key role in shaping the high-mass end of the BH mass distribution, while producing an excess at $35\Msun$. Both these studies demonstrate that SMT is an important formation channel for merging BBHs, and that mass transfer during the STAR+STAR phase is essential in determining what STAR+BH systems form in the first place.

When modeling BBH mergers, \citet{Bavera+23} were the first to also model the accretor in detail during the STAR+STAR phase, while also using detailed binary-star models in the STAR+BH phase. Most recently, \citet{Xu+25a} showed, with a set of detailed binary models at SMC metallicity, that mass accretion during the STAR+STAR phase can influence the subsequent evolution of the accretor, with possible implications for mass-transfer stability during the STAR+BH phase and the resulting BBH merger properties, highlighting the importance of the STAR+STAR phase.

Within the context of BBH mergers, the effect of the STAR+STAR phase has only just started to be explored, but current studies of the STAR+STAR interactions show that it is an important phase in the formation of BHs. \citet{Langer+20} explored the O/B+BH population using a grid of detailed stellar binaries and found a bi-modality in their period distribution originating from Case A and Case B mass transfer stability. 
The contribution of Case A mass transfer only became apparent when using detailed binary models due to issues in modeling Case A mass transfer in rapid population synthesis codes \citep[see for example][]{Romero-Shaw+23, Shikauchi+25}.
The work by \citet{Xu+25} extends the prediction of O/B+BH systems using detailed binary models at SMC metallicities, while the companion work \citet{Schurmann+25} highlights the differences between the rapid and detailed population synthesis approaches. Both find quantitative differences between the predicted O/B+BH populations from rapid and detailed population synthesis codes, further highlighting the need for detailed modeling of the STAR+STAR phase.

In this work, we present a full ZAMS-to-BBH-merger analysis using detailed binary models for each phase of mass transfer, focusing on the SMT channel. This work explores how the different evolutionary phases link together and aims at disentangling the important processes during each mass transfer phase in shaping the resulting BBH mergers.
We present the population synthesis model and physics choices in Section \ref{sec:method}. Section \ref{sec:BBH_merger_formation} describes the formation of BBH mergers at low-metallicity, showing how main-sequence (Case A) mass transfer is dominant in the formation of BBH mergers. Section \ref{sec:case_B} discusses why post-main-sequence (Case B) mass transfer does not produce BBH mergers within the Hubble time. Section \ref{sec:metallicity_dependence} expands the formation of BBH mergers to higher metallicities, showing how orbital widening during the STAR+BH phase limits BBH formation above $0.2\Zsun$. Section \ref{sec:natal_kicks} discuss the impact of the natal kick and in Section \ref{sec:discussion}, we discuss the impact of our physics assumptions. We summarize our conclusions in Section \ref{sec:conclusion}.

\section{Method} \label{sec:method} 

We use the binary population synthesis code \posydon v2.1.0 \citep{Fragos+23, Andrews+25} to create synthetic populations of BBH mergers at eight metallicities: $10^{-4}, 10^{-3}, 10^{-2}, 0.2, 0.45, 0.1, 1,$ and $2\Zsun$, with $\Zsun=0.0142$. \posydon uses grids of single- and binary-star models, computed using the 1D stellar evolution code \mesa \citep{Paxton+11, Paxton+13, Paxton+15, Paxton+18, Paxton+19,Jermyn+23} to evolve populations of single stars and binaries from ZAMS to, for example, BBH merger. These binary grids include self-consistent modeling of internal rotation and angular momentum transport in the stellar interior, between the two stars and the orbit of the binary. Here we focus our description of \posydon on aspects essential for BBH merger formation through SMT; for more specific details, see \citet{Fragos+23} and \citet{Andrews+25}.

For single star and detached (non-interacting) binary evolution, \posydon uses grids of single non-rotating stars that have been evolved up to white dwarf formation or carbon depletion. These either start at the ZAMS or as a helium star at core helium ignition, and we hereafter refer to them as \texttt{single-HMS} and \texttt{single-HeMS}, respectively. For potentially interacting binaries, there are three detailed binary-star model grid types in \posydon, corresponding to different evolutionary phases. For the STAR+STAR phase, there are the \HMSHMS grids, which cover the evolution from ZAMS until either unstable mass transfer occurs, a white dwarf is formed, or core carbon depletion is reached in either of the stars. For systems with a compact object (CO) companion, \posydon uses the \texttt{CO-HMS} and \texttt{CO-HeMS} grids depending on the properties of the stellar component in the binary. These three binary grids come in normal and \texttt{RLO} variants, where the grids start at Roche lobe overflow (RLO). For example, the \HMSHMS grid starts with two zero-age main-sequence binaries synchronized with the orbit. The \COHMSRLO grid, on the other hand, starts when the hydrogen-rich star overflows its Roche lobe for the first time. This allows \posydon to track the evolution of eccentric binaries until RLO occurs, after which the binary is circularized and matched to the \texttt{RLO} grids. These grid versions are created during the post-processing of the \mesa binary-star models, when the initial-final interpolators, used in this work, are also trained \citep[for more details, see][]{Fragos+23, Andrews+25}.

For hot, hydrogen-rich stars, \posydon uses the stellar wind mass loss from \citet{Vink+00} with a $(Z/Z_\odot)^{0.68}$ dependence on metallicity \citet{Vink+01}, while for helium-rich hot stars, the Wolf-Rayet-like winds from \citet{Nugis+00} are used. Furthermore, for stars reaching the Humphreys-Davidson limit, an enhanced mass loss of $10^{-4} \Msun\,\mathrm{yr}^{-1}$ is applied \citep{Belczynski+10a}.
\posydon uses an exponential core overshooting prescription, fitted to match the calibration from \citet{Brott+11}, without overshooting in shell burning and shell convective regions. Furthermore, an inefficient semi-convection parameter of $\alpha_\mathrm{sc}=0.1$ is chosen. 
For rotational mixing and internal angular momentum transport, \posydon follows the \texttt{MIST} project \citep{Choi+16}, and a more detail descriptions of the implementation of rotation and internal mixing processes can be found in section 3.2.3 in \citet{Fragos+23}.

When a star reaches carbon exhaustion, we use the delayed remnant mass prescription of \citet{Fryer+12} to determine the BH mass. The choice of prescription impacts the low mass regime of the BH formation, where the amount of fallback differs significantly per prescription. Above $M_\mathrm{CO{-}core} \simgt 11 \Msun$, nearly all prescriptions assume direct-collapse of the cores into the BH, and the difference between prescriptions is minimal. We explore other remnant mass prescription in Appendix \ref{app:other_SN_prescriptions} to show its impact on the mass and mass ratio distributions. The spin of the resulting BH is estimated based on \citet[][see their appendix D]{Bavera+21}, where the collapse of the stellar profile at core carbon depletion and accretion onto a proto-BH of $2.5\Msun$ is followed, including the angular momentum of the infalling matter \citep[see also Section 8.3.4 in][]{Fragos+23}.

In our default model, we assume no natal kick for a clearer linking between the binary physics and outcome of the STAR+STAR phase and the beginning of the STAR+BH phase. With no kicks, a small eccentricity up to 0.2 can still be gained from the instantaneous mass loss during the supernova \citep{Blaauw+61}. In Section \ref{sec:natal_kicks}, we implement a one-over-mass scaled kick drawn from a Maxwellian distribution with $\sigma=265$\,km\,s$^{-1}$ \citep{Hobbs+05} to show the impact a kick prescription has on the SMT channel properties\footnote{Although \citet{Disberg+25} propose a correction to the \citet{Hobbs+05} natal kick distribution, we use the Maxwellian distribution proposed by \citet{Hobbs+05}, since it remains widely adopted in the literature to date.}. We draw the kick angles and anomaly from uniform distributions, and account for the effect of the natal kick on the tilt of the orbit by following \citet{Kalogera+96a} and \citet{Wong+12}. Because we do not realign the BH spin axis during mass transfer, we calculate a combined tilt after the second supernova \citep[also see section 8.3.5 in][]{Fragos+23}. These final tilts are used in the calculation of \chieff for the BBH mergers. For the pair-instability supernova regime, we adapt the prescriptions by \citet{Hendriks+23} to fit the peak and PISN limit of \citet{Farag+22}, which leads to a $\Delta M_\mathrm{CO{-}core}=0\Msun$ and $\Delta M_\mathrm{PPI} = -20\Msun$ \citep[For more details, see section 2.2.1 in][]{Andrews+25}.

For mass transfer between two stars, \posydon uses the \texttt{contact} scheme for RLO on the main-sequence, allowing for both stars to fill their Roche lobes simultaneously and a contact system to form. Post main-sequence, \posydon switches to the \texttt{kolb} mass transfer scheme \citep{Kolb+90}, which allows the star to expand past its Roche lobe. Since we self-consistently model the mass transfer and detachment, donor stars can end their mass transfer well before being fully stripped to their core, which is especially true in sub-solar metallicity binaries \citep[e.g. ][]{Klencki+20}. We limit accretion onto a BH to the Eddington limit and assume angular momentum accretion as described in section 4.2.3 in \citet{Fragos+23}. For a stellar companion, we follow \citet{deMink+13} for the specific angular momentum carried by the accreted material, which in most cases rapidly spins up the accretor to near critical rotation. To keep the accretor below the critical rotation threshold, we assume rotational-limited accretion, where the increase in stellar rotation boosts the mass loss through stellar winds. Effectively, this leads to efficient mass transfer if a mechanism exists that spins down the accretor, and inefficient mass transfer if the spin-down is inefficient. One such mechanism is tides, which can cause a rapidly rotating star to spin down when strong enough. In close orbits of massive stars, this is particularly true, even during mass transfer \citep{PortegiesZwart+96, Hurley+02}. \posydon follows the tides as implemented by \citet{Hut+81}, as described in Section 4.1 in \citet{Fragos+23}. Since no mass transfer efficiency is set a priori, short-period binaries have more conservative mass transfer than wider systems (see appendix F in \citealt{Sen+22} for the same effect, and see \citealt{Rocha+24} and \citealt{Zapartas+25} for examples of the mass transfer efficiency within \posydon).

The stability of mass transfer in \posydon is determined self-consistently for each \mesa binary model. Mass transfer is considered unstable if the mass transfer rate exceeds either 0.1\Msun yr$^{-1}$ or the photon trapping radius for compact object accretors \citep{Begelman+79}. In addition, systems experiencing outflow through the $L_2$ point are classified as unstable, as material lost through this point removes substantial angular momentum from the binary, leading to a rapid orbital inspiral and the onset of a CE phase \citep{Tylenda+11, Nandez+14}. For post-main-sequence stars, the $L_2$ radius is determined following the prescription from \citet{Misra+20}, while the definition from \citet{Marchant+16} is used for main-sequence over-contact binaries. Finally, the mass transfer is considered unstable if stars are in contact, and one of the stars has evolved off the main-sequence \citep[See Section 4.2.3 in][for a more detailed description of the stability criteria]{Fragos+23}. If one of the conditions is reached, the \mesa model is stopped and tagged with an appropriate termination flag.

We sample $10^6$ binaries per metallicity from a \citet{Kroupa+01} initial mass function from $7\Msun$ to $270\Msun$ and flat mass ratio distribution, while the separation is sampled from log uniform between $5\Rsun$ and $10^5\Rsun$. We account for the unsampled parameter space by reweighting the binaries accordingly, where we assume a binary fraction of 70\%.
Throughout this work, we focus exclusively on BBH mergers formed through the SMT channel, where systems undergo SMT during both the STAR+STAR and STAR+BH phases. References to "BBH mergers" in this work consider only this formation channel.
We classify mass transfer episodes by the evolutionary stage of the donor star at the first mass transfer in each phase. For example, Case A refers to mass transfer initiated while the donor is on the main-sequence, while Case B refers to mass transfer once it has evolved off the main-sequence up to core-helium depletion. At lower metallicities, systems commonly experience multiple mass transfer episodes within a single phase due to partial envelope stripping and re-expansion \citep{Gotberg+17, Laplace+20}. While these interactions are modeled in the \mesa models, our mass transfer classification only depends on the initial mass transfer during that phase.

Figure \ref{fig:example_binary} shows how a BBH merger progenitor evolves through the \posydon grids with each phase marked in orange. The STAR+STAR phase is completely modeled with the \HMSHMS grids in \posydon for BBH mergers in the SMT channel. The binaries in this grid start as two ZAMS synchronized stars over a large range of orbital periods and mass ratios. After the formation of the first BH, most systems evolve through a detached phase, where the secondary is matched to a single star and the eccentric binary is evolved until it reaches carbon depletion or fills its Roche lobe. At RLO, the binary enters the \COHMSRLO grid and either reaches core carbon depletion or one of the unstable mass transfer criteria in \posydon. This means that for BBH mergers formed through the SMT channel, all the binary interactions are directly simulated within our grids of detailed models. The only evolutionary phases that are not modeled with \mesa binary simulations are the supernova explosion and detached phases.

We define $M_1$ as the more massive star at ZAMS (primary), and $M_2$ as the less massive star (secondary). We use the term "donor star" ($M_\mathrm{donor}$) to refer to the component losing mass during the mass transfer. Generally, this corresponds to $M_1$ during the STAR+STAR phase and $M_2$ during the STAR+BH phase. At BBH merger, we use $M_1$ ($M_2$) for the BH originating from the initially more (less) massive star. We use $M_\mathrm{max}$ ($M_\mathrm{min}$) for the more (less) massive BH, independent of which star formed it, which can occur due to mass ratio reversal, where the initially less massive star produces the more massive BH in the merging BBH.

\section{Formation of BBH mergers at low metallicity} \label{sec:BBH_merger_formation}

First, we analyze the population of BBH mergers at $10^{-4}\Zsun$. Although this metallicity is not representative of the bulk of cosmic star formation, we find that it is where the SMT channel is most efficient in producing BBH mergers within the Hubble time. We begin, in Section \ref{sec:example_model}, with a representative example model to illustrate the typical evolution in the SMT channel leading to BBH mergers. Afterwards, we discuss the STAR+BH phase in Section \ref{subsec:STAR_BH_phase}, which is modeled in the \COHMSRLO grid. We start our exploration with the STAR+BH grid, since it is closest to BBH formation and has been explored in previous works \citep{Marchant+21, Gallegos-Garcia+21, Klencki+26}. Finally, Section \ref{subsec:STAR+STAR} discusses the effect the STAR+STAR phase has on BBH merger formation.

\subsection{Representative example model} \label{sec:example_model}

\begin{figure*}
    \centering
    \includegraphics[width=\linewidth]{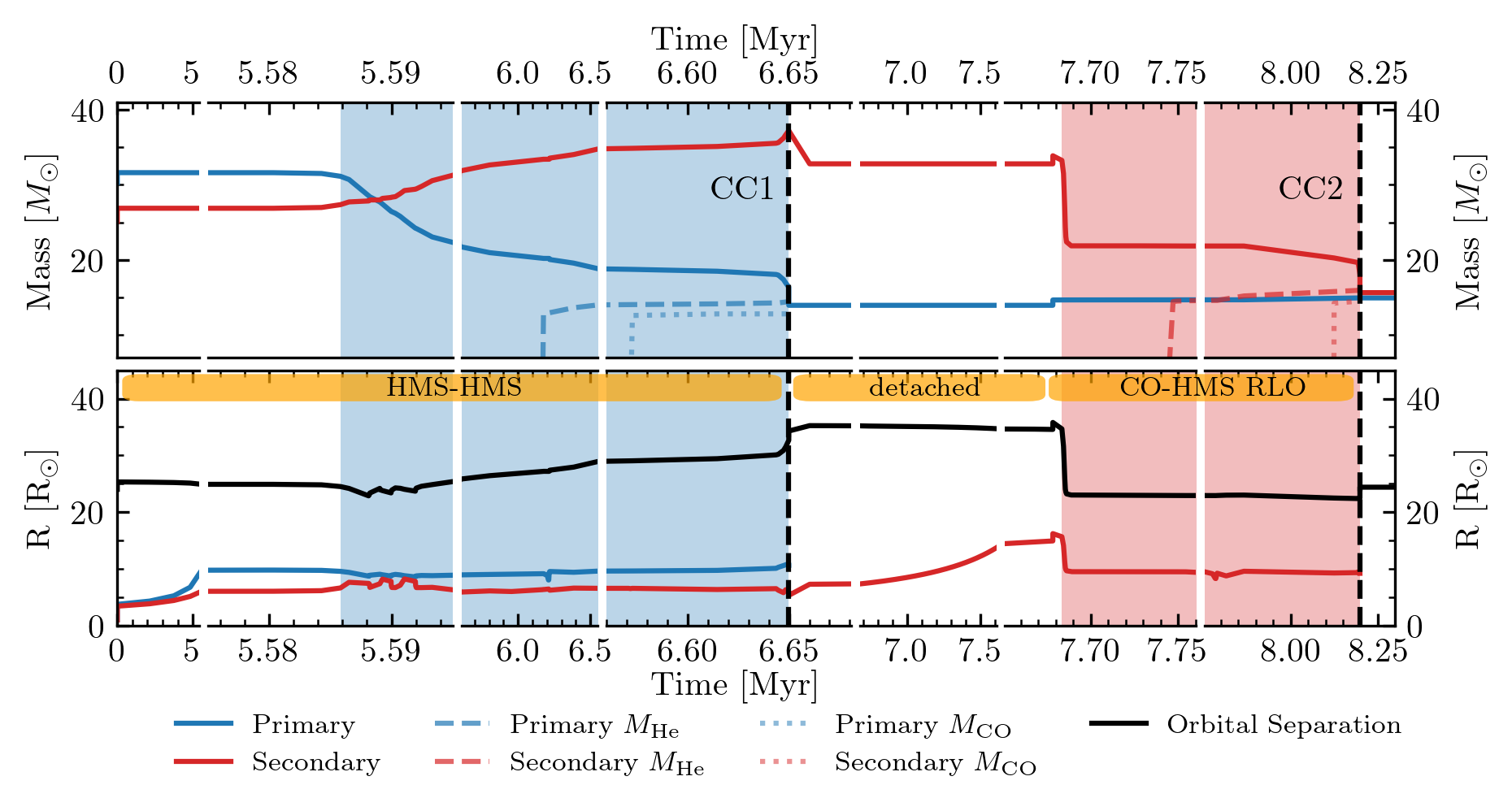}
    \caption{Representative binary from the SMT channel leading to a BBH merger at $10^{-4}\Zsun$ with $M_1=32\Msun$, $M_2=27\Msun$, and $P=1.9$ days at ZAMS. The time series is divided into different scales to show details in the evolution on shorter timescales. The top row shows the mass evolution of the primary (blue) and secondary (red), while the bottom row shows their stellar radius evolution. The shaded regions are when RLO occurs, colored according to the donor star. The black solid line tracks the orbital separation, and the vertical dashed lines mark the times of core collapse for each component. Between 5.58 Myr and 5.59 Myr, the first mass transfer phase starts. After a fast thermal timescale interaction, the mass ratio flips, and the mass transfer continues on a nuclear timescale. At ${\sim}7.68$ Myr, the detached binary with an eccentricity of 0.001 is circularized and matched to the nearest \COHMSRLO grid model, causing the slight increase in mass, period, and radius. The steep drop afterwards is caused by thermal timescale mass transfer. The small changes in radii at $6.2$ Myr and $8.05$ Myr are from readjustment after core hydrogen depletion in the primary and secondary, respectively.
    In the top row, the He-core ($M_\mathrm{He}$) and CO-core ($M_\mathrm{CO}$) masses are shown with dashed and dotted lines, respectively. This system is evolved with the same simulation setup as the population models, but the nearest-neighbor method (matching to the closest down-sampled precomputed track) is used instead of initial–final interpolation to show the evolution over time \citep[for more details, see][]{Fragos+23}. The orange bars indicate the \posydon step or grid that is used for the evolution of the binary.}
    \label{fig:example_binary}
\end{figure*}

To provide a reference point for further analysis, we show the evolution of a representative BBH merger progenitor in Figure \ref{fig:example_binary} (see Appendix~\ref{app:general_population_@_10^-4} for the general population properties). At ZAMS, the system has $M_1=32\Msun$, $q=0.85$ ($M_2=27\Msun$) in a tight 1.9-day orbit. After ${\sim}5.6$ Myr, the primary fills its Roche lobe while on the main-sequence, a Case A mass transfer.
Following a rapid initial phase that reverses the mass ratio, the binary enters a nuclear-timescale mass transfer phase that continues until the primary reaches core carbon depletion, which is similar to the interactions explored in detail by \citet{Sen+22} in the context of Algols. The phase from ZAMS to core carbon depletion is completely modeled within the \HMSHMS grid of \posydon.
The tight orbital configuration allows tides to spin down the accretor, resulting in a high accretion efficiency, mitigating the effects of rotation-limited accretion.

At core-collapse, the primary becomes a $14\Msun$ BH, while the secondary has increased to $37\Msun$ due to the preceding mass transfer, which also increased the orbital period to 3.2 days. After a phase of detached evolution, the secondary also fills its Roche lobe on the main-sequence with a mass of $34\Msun$ at 7.68 Myr, initiating another Case A mass transfer that lasts until core carbon depletion. Because the detached system is matched to the \COHMSRLO grid, a small increase in mass, separation, and radius can be seen at 7.68 Myr, followed by a steep drop in the same properties caused by an initial fast, thermal-timescale phase of Roche lobe overflow. Despite being Eddington-limited, the BH accretes approximately $1\Msun$ during the mass transfer, producing a final primary BH mass of $14.9\Msun$. The secondary has been stripped to $18\Msun$ and collapses to a $15.6\Msun$ BH. This mass transfer phase is modeled within the \COHMSRLO grid, which starts at RLO and ends at core carbon depletion for SMT systems. The mass ratio reversal, where the secondary produces the more massive BH, occurs in 51\% of BBH mergers at $10^{-4}\Zsun$. The final BBH system merges ${\sim}7.2$ Gyr after ZAMS. The tight initial orbital configuration is required to produce a BBH system with a merger time within the Hubble time and is representative of the general population of BBH mergers.

\subsection{The STAR+BH phase} \label{subsec:STAR_BH_phase}

\begin{figure*}
    \centering
    \includegraphics[width=\linewidth]{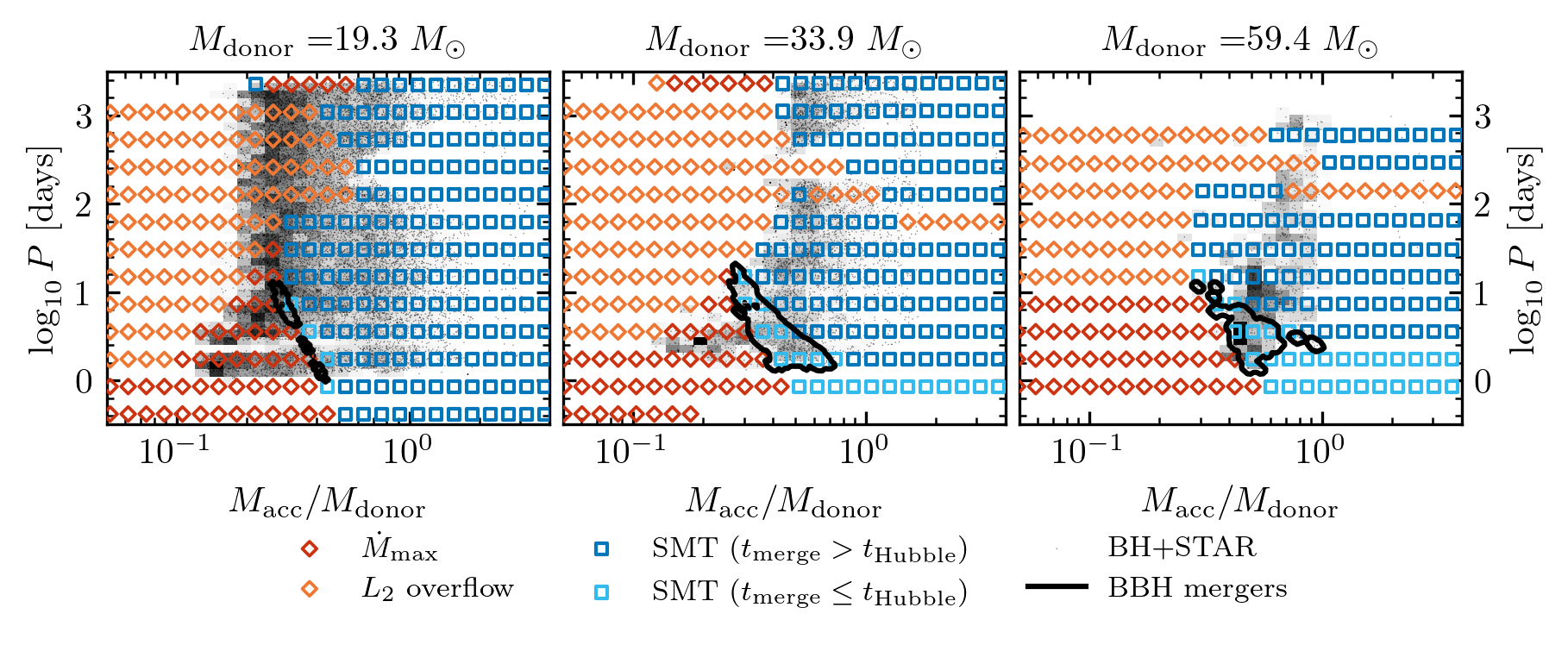}
    \caption{Example grid slices at $10^{-4}\Zsun$ for fixed $\Mdonor = 19.3\Msun$ (left), $33.9\Msun$ (middle), and $59.4\Msun$ (right) for the \COHMSRLO{} grid, where the models start at RLO. The diamonds indicate unstable mass transfer with $L_2$ overflow (orange) and $\dot{M}_\mathrm{max}$ (red) separated. Stable mass transfer is indicated with squares with mergers (light blue) and non-mergers (dark blue) within the Hubble time based on the same remnant mass calculation as our default population. The STAR+BH systems evolved in a \posydon population with $\Delta M \pm 2$ around $M_\mathrm{donor}$ going into this grid slice are shown as a 2D histogram behind the grid markers, as well as dots for the exact systems. The darker the shading, the higher the density of systems. The black contour contains 100\% of the STAR+BH system that produces BBH mergers, showing the overlap between the evolved population and region that leads to BBH mergers in the \COHMSRLO{} grid. The latter can be slightly off from the exact location of progenitor properties due to underlying histogram binning, but also due to the 4-dimensional nature of the grids. It provides an indicator of the parameter space region of the BBH mergers progenitors.}
    \label{fig:example_Mdonor}
\end{figure*}

The example model (Section~\ref{sec:example_model}) and the general population (Appendix~\ref{app:general_population_@_10^-4}) demonstrate that tight initial orbits are crucial for producing BBH mergers within the Hubble time through SMT, implying that most BBH progenitors will predominantly undergo Case A mass transfer as their first interaction. Additional later interactions can occur in the same phase, for example, the example model in Figure \ref{fig:example_binary} undergoes Case A, Case B, and Case C mass transfer, as one continuous nuclear timescale interaction during the STAR+STAR phase and during the STAR+BH phase, as the shaded regions highlight. To understand how these initial conditions lead to BBH mergers, we examine the locations in the $P{-}q$ parameter space of progenitors within the \posydon grids and identify the key physical processes shaping their evolution.

Figure \ref{fig:example_Mdonor} shows representative 2D slices of the \COHMSRLO grid for donor masses $19.3\Msun$, $33.9\Msun$, and $59.4\Msun$\footnote{See Figure \ref{fig:MBH_grid_slice} for the other representation of this grid where $\MBH$ is fixed.}.
The colored markers are the outcomes of the \COHMSRLO grid, while the grayscale dots and shading are the STAR+BH binaries, whose outcome will be determined by the \COHMSRLO grid. The grayscale shading shows the density of the systems in a 2D histogram of period and $\MBH/\Mdonor$ mass ratio of the full metallicity-specific population. The location of the grayscale dots is determined by the previous evolutionary phases (STAR+STAR and supernova). The population systems leading to BBH mergers within a Hubble time through SMT are delineated by the black contour.

\subsubsection{\posydon \COHMSRLO{} grid}

For a STAR+BH system to form a merging BBH within the Hubble time, SMT during the STAR+BH phase needs to shrink the orbit sufficiently for gravitational wave emission to lead to contact between the two BHs. The light blue squares in Figure \ref{fig:example_Mdonor} shows the binary-star models in the \COHMSRLO grid producing a BBH merger within the Hubble time ($t_\mathrm{merger} \leq t_\mathrm{Hubble}$). The blue square markers are binary-star models experiencing SMT and not leading to a merger within $t_\mathrm{Hubble}$, which is the majority of SMT models. Only systems with mass ratios that narrowly avoid instability or already have short periods at the onset of RLO ($P\sim1$ day) successfully produce BBH mergers. This narrow region for BBH merger formation in the grid resembles other grids available in the literature \citep{Marchant+21, Gallegos-Garcia+21, Klencki+26}. In \posydon it is assumed that $L_2$ overflow leads to dynamically unstable mass-transfer (orange diamond markers), resulting in most post-main-sequence interactions becoming unstable\footnote{The large radial expansion post-main-sequence comes from an interplay between inefficient semi-convection causing a large radial expansion after core-hydrogen depletion and switching to the \texttt{kolb} mass transfer scheme after the main-sequence to allow for radial expansion past the Roche lobe.}. Since this differs from other STAR+BH grids in the literature, we discuss the effect of this choice on the BBH merger through SMT in detail in Section \ref{sec:case_B}. Generally, though, the parameter space of the STAR+BH grids in the literature, where their binary-star models lead to a BBH merger through SMT with $L_2$ outflow (low $\MBH/\Mdonor$ and large $P$) is sparsely populated in \posydon. This is due to inefficient Case B mass transfer in the preceding STAR+STAR phase, preventing systems from reaching this region, limiting the impact of the $L_2$ outflow stability treatment on the formation of BBH through the SMT channel.

\subsubsection{STAR+BH population systems}

The systems entering the \COHMSRLO{} grid in the middle panel with $\Mdonor = 33.9\Msun$ in Figure \ref{fig:example_Mdonor} (the grayscale dots and shading) can be split into two groups based on orbital period at RLO: Below $P<10$ days, binary systems span mass ratios from $0.1{-}2$, while above $P>10$ days the \MBH/\Mdonor mass ratios only extend down to ${\sim}0.4$\footnote{The upper-limit is set by the primary experiencing pair-instability and, thus, not producing a BH above ${\sim}60\Msun$, which translates to $q\approx2$ for $\Mdonor = 33.9\Msun$ in the middle panel of Figure \ref{fig:example_Mdonor}.}. The wide range of mass ratios below $P<10$ days is due to a high efficiency in accretion during Case A mass transfer in the STAR+STAR phase. This mass transfer allows the secondary (\Mdonor in the CO-HMS grid) to gain a significant amount of mass, while the primary forms a lower mass BH compared to isolated evolution \citep{Schurmann+24}, producing low $\MBH/\Mdonor$ ratios. A similar effect can be seen in the example model in Figure \ref{fig:example_binary}, where the secondary gains ${\sim}10\Msun$ during the STAR+STAR phase, producing a mass ratio of $\MBH/\Mdonor\sim 0.4$. Due to the example being selected to produce a BBH merger, its $\MBH/\Mdonor$ ratio is not as small as it could be.

In contrast, population systems with $P\simgt10$ days have a more restrictive minimum mass ratio of $\MBH/\Mdonor\simgt 0.4$. These wider systems underwent Case B mass transfer during the STAR+STAR phase, where the mass of the helium core is already set when the interaction occurs. 
At $10^{-4}\Zsun$, the donor star retains a large hydrogen envelope \citep{Gotberg+17, Klencki+20} and its helium core mass is not altered by the interaction (for more details, See Section \ref{sec:case_B}). With the \citet{Fryer+12} delayed SN prescription, the mass of the helium core determines the mass of the BH, which is generally the case for most supernova prescriptions at high core masses. Therefore, the BH mass is effectively fixed, despite Case B or Case C mass transfer. Moreover, the partial-stripping reduces the amount of mass involved in the mass transfer, while at wider orbits, the mass gain of the companion is limited, as tides do not mitigate the rotationally-limited accretion anymore. The combined effects produce a higher minimum $\MBH/\Mdonor$ ratio than in Case A systems.

\subsubsection{BBH mergers}

The black contours in Figure \ref{fig:example_Mdonor} delineate the population systems that successfully produce BBH mergers within the Hubble time through SMT, which closely aligns with the light blue grid models, confirming that this region is responsible for BBH merger formation in \posydon. However, the shortest-period grid models ($P\sim 1$ day at RLO) are notably absent from the population. Despite these grid models yielding BBH mergers regardless of \MBH, such tight initial orbits in the STAR+STAR phase produce main-sequence contact binaries that undergo unstable mass transfer due to $L_2$ outflow (Section \ref{subsec:STAR+STAR}). As a result, this region of parameter space in CO-HMS is inaccessible to the SMT channel due to the STAR-STAR phase.

\subsubsection{Dependence on $\Mdonor$}

The left panel in Figure \ref{fig:example_Mdonor} with $\Mdonor=19.3\Msun$ shows that the SMT-to-BBH-merger grid region, indicated by the light blue squares, has shrunk. This implies that, for mergers to occur within the Hubble time, the binary must have mass ratios closer to the instability boundary.
For a less massive \Mdonor, the resulting second-born BH is also less massive, and thus, more efficient orbital shrinkage during the mass transfer is required compared to binaries with more massive components in order to merge within the Hubble time. As a result, only binaries with unequal mass ratios can undergo sufficient orbital tightening during the SMT phase, which has to become more unequal as $\Mdonor$ decreases. Eventually, the $\Mdonor$ becomes too low to form a BH, preventing the formation of a BBH merger below $\Mdonor \simlt 14\Msun$.

For the $\Mdonor=59.4\Msun$ in the right panel of Figure \ref{fig:example_Mdonor}, the outcomes of our \COHMSRLO grid are broadly similar to $\Mdonor=33.9\Msun$ with the parameter space of light blue SMT-to-BBH-merger models expanding. More massive BBH systems lose more angular momentum through gravitational wave emission, reducing the orbital shrinkage required from SMT to produce a merger within the Hubble time. Therefore, the grid models are showing an expanded parameter space that leads to a BBH merger through SMT at higher \Mdonor. The light blue squares, representing BBH mergers through SMT models, now extend to more equal mass ratios ($\MBH/\Mdonor$) for a wider range of orbital periods.
Moreover, the systems that merge independently of \MBH have expanded to wider initial periods ($P\sim 2$ days).
However, two population-level effects limit binary systems, marked in grayscale, from populating this expanded BBH merger grid parameter space. First, the $P\simlt1$ day regime remains unpopulated and the boundary shifts to slightly longer periods (${\sim}1.4$ days) compared to lower donor masses ($P\sim1$ day in $\Mdonor = 19.3\Msun$ in the left panel of Figure \ref{fig:example_Mdonor}). This happens because more massive stars have larger radii, making the contact binaries during the STAR+STAR phase more prone to instability (Section \ref{subsec:STAR+STAR}).
Secondly, the widest-period, lowest-\MBH/\Mdonor grid models that produce BBH mergers through SMT are unpopulated at high \Mdonor. This is likely because direct collapse of the primary creates a more massive BH, while a larger helium core fraction at higher mass limits the mass the secondary can gain during Case B, resulting in higher mass ratios.

\subsection{The STAR+STAR phase} \label{subsec:STAR+STAR}

\begin{figure*}
    \centering
    \includegraphics[width=\linewidth]{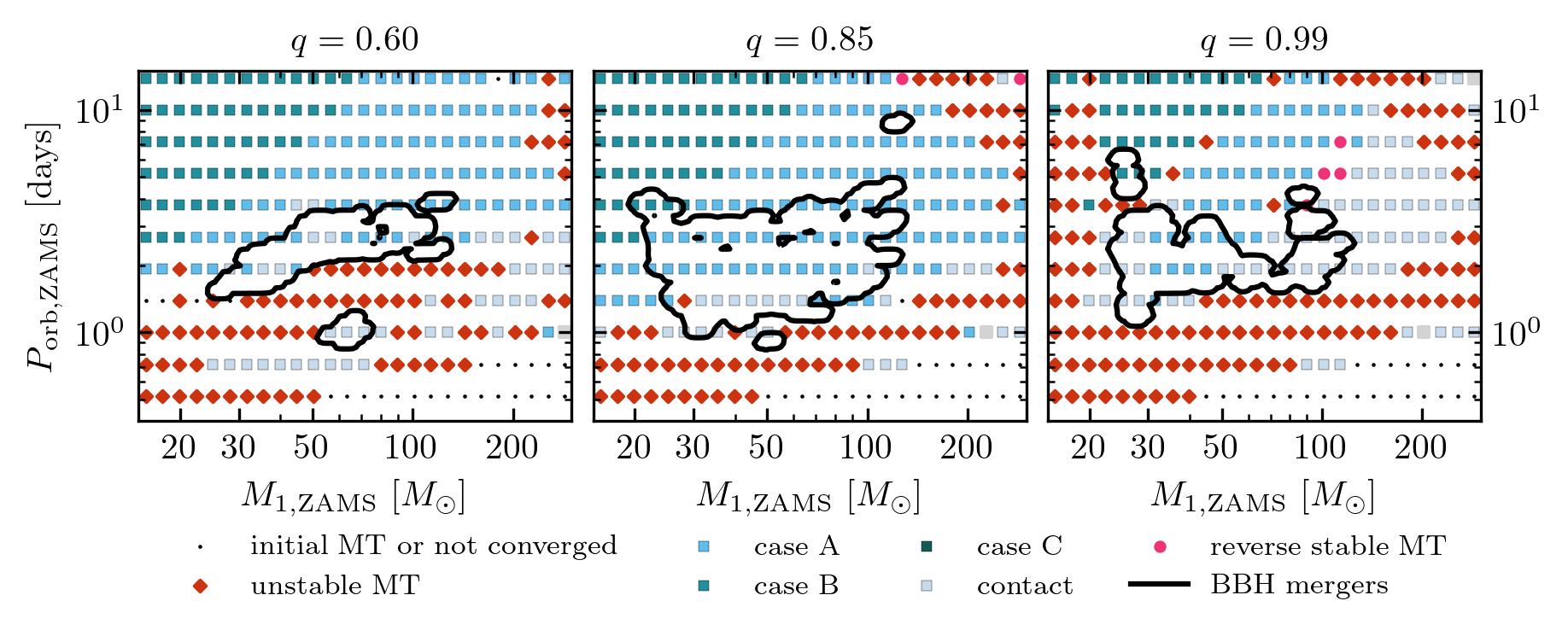}
    \caption{Grid slices of the HMS-HMS grid at $q=0.60$, $0.85$, and $0.99$ for $Z=10^{-4}\Zsun$. We do not show the grid below $M_1 < 15 \Msun$ because such systems do not contribute to the BBH merger rate. The non-converged (or initial RLOF), unstable, and reverse stable mass transfer \citep[see][ for reverse stable mass transfer models]{Briel+25} models are marked with black dots, red diamonds, and pink circles, respectively. Models without mass transfer are marked with gray squares. SMT systems are marked with four different colored squares, where the exact color indicates a contact phase during the MS (gray squares), Case A (light blue), Case B (teal), or Case C mass transfer (dark green) as the first interaction of the binary. We show the ZAMS properties of BBH merger progenitors by encircling all of them with a black contour, based on a smoothed 2D histogram.}
    \label{fig:HMS-HMS-example-grid}
\end{figure*}

The population of systems that produce BBH mergers is shaped by the interplay between the \COHMSRLO grids (as discussed in Section \ref{subsec:STAR_BH_phase} and the initial conditions for the STAR+BH phase inherited from the preceding evolutionary phases. As shown in Figure \ref{fig:example_Mdonor}, systems entering the STAR+BH phase (grayscale dots) only occupy a subset of the SMT parameter space producing BBH mergers within the Hubble time (light blue squares). We examine in this Section the population-level constraints originating from the STAR+STAR phase that determine the initial conditions of systems entering the STAR+BH phase (grayscale dots in Figure \ref{fig:example_Mdonor}).
In the representative \HMSHMS 2D grid slices in Figure \ref{fig:HMS-HMS-example-grid}, the black contour denotes the region of parameter space producing BBH mergers through SMT. In primary mass, this region is limited by the minimum BH mass of $2.5\Msun$ and the maximum stellar mass below the pair-instability supernova limit \citep[e.g,][]{Farmer+19, Renzo+22, Farag+22}. The binaries within the black contour nearly all have tight initial orbits with $P\simlt 4$ days, undergoing Case A mass transfer as their first interaction, similar to the example model in Figure \ref{fig:example_binary}. The short $P_\mathrm{ZAMS}$ are a general feature for BBH merger progenitors in the SMT channel. Additionally, we find a lower period boundary set by the stability of main-sequence mass transfer, and an upper period boundary by the Hubble merger time regime in the \COHMSRLO grid. We will first discuss the lower period boundary and then discuss how the \HMSHMS grid links to the initial properties of the STAR+BH phase.

\subsubsection{Minimum initial period boundary}

At all three mass ratios in Figure \ref{fig:HMS-HMS-example-grid}, the minimum period of the BBH progenitor regime (black contour) extends down to periods approaching the unstable mass transfer regime (red diamonds). These very short-period binaries reach $L_2$ outflow during a contact phase, which triggers dynamical instability and CE evolution in \posydon. The lower boundary for BBH mergers shifts to longer periods with decreasing mass ratio, as can be seen when comparing the left and middle panels in Figure \ref{fig:HMS-HMS-example-grid}. For example, for $M_{1,\mathrm{ZAMS}}\sim50\Msun$, the first stable model producing a BBH merger is at $P_\mathrm{ZAMS}\sim 1$ day for $q=0.85$ and $P_\mathrm{ZAMS}\sim1.8$ day for $q=0.60$. Because the mass transfer phase with $M_1>M_2$ lasts longer in the $q=0.60$ model (a larger fraction of the donors mass is transferred), this system can reach a smaller minimum orbital separation than the $q=0.85$ system during the mass transfer. As a result, the $L_2$ instability criterion is reached more readily in $q=0.60$ systems, and systems with longer initial periods can avoid the instability at lower mass ratios.

When the mass ratio approaches close to unity, the minimum period boundary for SMT increases again, as can be seen when comparing the red diamonds in the middle and right panels in Figure \ref{fig:HMS-HMS-example-grid}. For the $M_1\sim 50\Msun$ at $q=0.99$, the minimum stable period is at $P_\mathrm{ZAMS}\sim1.6$ day, slightly longer than at $q=0.85$. This is due to the companions having a larger radius and more similar main-sequence lifetimes when the mass ratio is near unity. This leads to binaries entering a contact phase that lasts until the end of core hydrogen burning. Following this, the rapid post-main-sequence expansion triggers CE evolution in \posydon. The mass ratio dependence of this instability, combined with the $L_2$ outflow instability, leads to the shortest stable $P_\mathrm{ZAMS}$ around $q_\mathrm{ZAMS}\sim0.75{-}0.8$.

For all mass ratios, as $M_1$ increases, the radii of both binary components on the main-sequence increase, and longer $P_\mathrm{ZAMS}$ are required to avoid unstable mass transfer on the main-sequence. This leads to a slight upward trend in the minimum $P_\mathrm{ZAMS}$ for BBH merger formation as $M_1$ increases, which is visible in all three panels in Figure \ref{fig:HMS-HMS-example-grid}. This trend is the cause of the increase in the minimum period of the STAR+BH population as a function of \Mdonor, which is visible in Figure \ref{fig:example_Mdonor} as grayscale dots shifting from $P\sim1$ day to $P\sim1.6$ day between $\Mdonor=19.3\Msun$ and $\Mdonor=59.4\Msun$.
Additionally, in the middle panel of Figure \ref{fig:HMS-HMS-example-grid}, for systems with $M_1\simgt100\Msun$, the tightest stable orbital configurations near the PISN limit no longer lead to BBH mergers, i.e., the black contour moves away from the red diamonds to longer periods. The accretion efficiency in the tightest stable orbital periods is near 100\%, leading to the secondary being shifted into the PISN regime (the region between the black contour and the red diamonds). This restricts the tightest SMT systems near the PISN limit from forming BBH mergers. Due to reduced strength of spin-down via tides, a longer-period binary has a lower mass accretion efficiency, and therefore the secondary avoids entering the PISN regime. This sets a slightly more restrictive lower period bound than the mass transfer stability criteria at the highest $M_\mathrm{1,ZAMS}$.

\begin{figure}
    \centering
    \includegraphics[width=\linewidth]{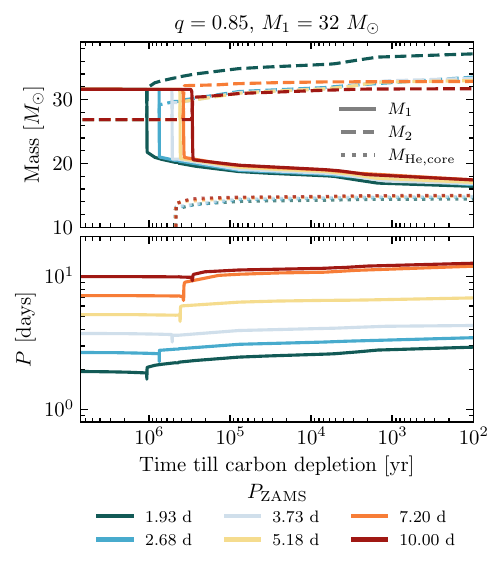}
    \caption{Evolution of $M_1=32\Msun$ and $q=0.85$ example binaries with different $P_\mathrm{ZAMS}$, as a function of time till carbon depletion. The top panel shows the mass evolution for the primary and secondary, while the bottom panel shows the period evolution. Similar to the model in Figure \ref{fig:example_binary}, the mass transfer starts with a rapid phase, which shrinks the orbit, causing the dips in the period evolutions. However, the mass ratio flips, and the orbit widens again. In general, a tighter initial orbit leads to a shorter period and a more massive companion at carbon depletion, while the helium core mass is independent of $P_\mathrm{ZAMS}$. The STAR+BH conditions of these binaries are marked with a star in Figure \ref{fig:MBH_grid_slice}.}
    \label{fig:period_dependence}
\end{figure}

\begin{figure}
    \centering
    \includegraphics[width=\linewidth]{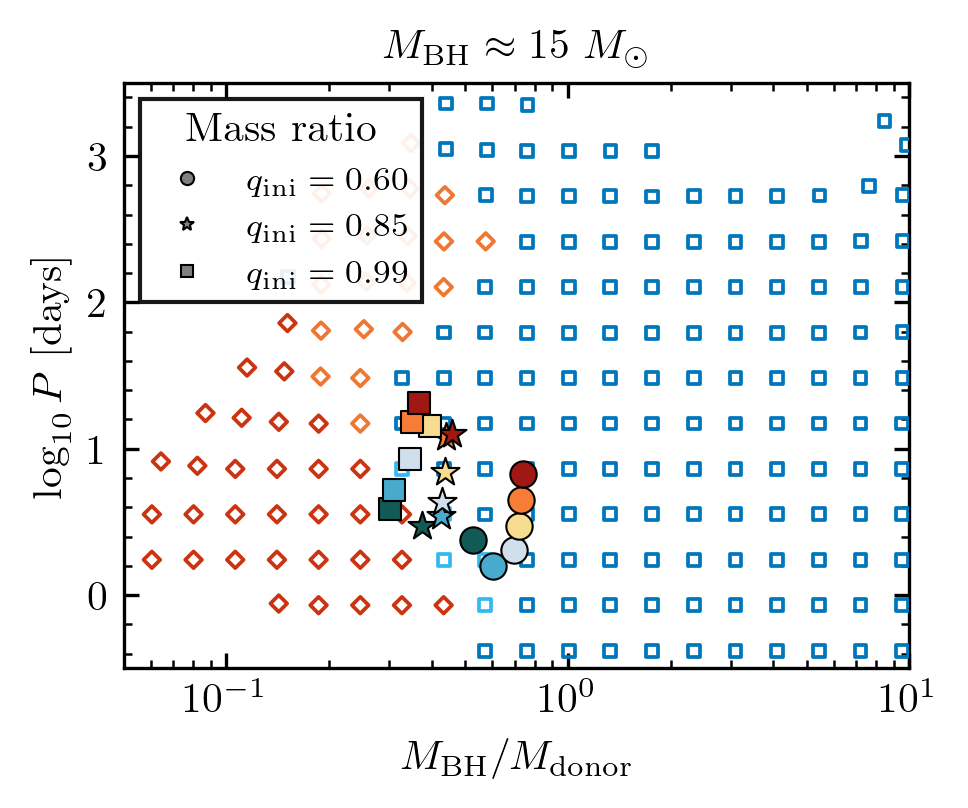}
    \caption{\COHMSRLO{} grid slice for a fixed $\MBH\approx15\Msun$, which is the remnant mass of the primary in the example models ($M_\mathrm{1,ZAMS}\approx 32\Msun$) from Figure \ref{fig:period_dependence}. The star markers are the properties of the example models with $q=0.85$ after the first supernova for different initial periods, where the coloring is the same as in Figure \ref{fig:period_dependence}. The squares and circles are models with the same $P_\mathrm{ZAMS}$ and $M_\mathrm{1, ZAMS}$ but with $q=0.99$ and $q=0.60$, respectively. The markers and colors of the models from the \COHMSRLO{} grid are the same as in Figure \ref{fig:example_Mdonor}, but here \MBH is fixed instead of \Mdonor.  As $P_\mathrm{ZAMS}$ increases, the STAR+BH binaries move outside the SMT BBH merger regime.}
    \label{fig:MBH_grid_slice}
\end{figure}

\subsubsection{Maximum period boundary}

The maximum $P_\mathrm{ZAMS}$ is determined by SMT models in the \COHMSRLO grid that merge within the Hubble time. To link the \HMSHMS period to the \COHMSRLO grid, Figure \ref{fig:period_dependence} shows the \HMSHMS evolution of the example binary from Figure \ref{fig:example_binary} and five longer period systems with the same total mass and mass ratio. The top panel shows that independent of the initial period, the final core masses are within 1\Msun, which leads to similar BH masses of ${\sim}15\Msun$ in the STAR+BH phase. Thus, we show a representation of the \COHMSRLO grid with a fixed \MBH in Figure \ref{fig:MBH_grid_slice}. The locations of the example models are marked with stars in the same colors as in Figure \ref{fig:period_dependence}. Because the companion in short-period systems accretes more efficiently due to stronger tidal interactions, \Mdonor increases and \MBH/\Mdonor decreases, resulting in the right shift of the models in Figure \ref{fig:MBH_grid_slice} with increasing $P_\mathrm{ZAMS}$. At the same time, initially longer-period systems reach carbon depletion with a longer final period. This is primarily due to mass ratio reversal during the mass transfer, which counteracts the initial orbital shrinkage, as the period evolution in Figure \ref{fig:period_dependence} shows. Any additional mass loss from the primary will further widen the orbit. As a result, the rightward shift is combined with an upward shift in Figure \ref{fig:MBH_grid_slice}, moving an initially longer-period binary outside the region that leads to mergers within a Hubble time in the \COHMSRLO grid, setting a maximum period in $P_\mathrm{ZAMS}$. The same effect is present for the other initial $q$ values shown in Figure \ref{fig:MBH_grid_slice}. In summary, although different $P_\mathrm{ZAMS}$ produce similar first-born BH masses, reduced accretion efficiency onto the stellar companion and increased orbital widening shifts the STAR+BH systems to configurations unable to merge within the Hubble time, thereby setting a maximum $P_\mathrm{ZAMS}$.

Because models with higher $q_\mathrm{ZAMS}$ enter the \COHMSRLO grid slice in Figure \ref{fig:MBH_grid_slice} with wider orbital periods, and due to the SMT-to-BBH-merger model region (light blue squares) shrinking with increasing \Mdonor, the maximum $P_\mathrm{ZAMS}$ that produces BBH merger is lower for $q=0.99$ than for $q=0.85$, as can be seen in the right and middle panels of Figure \ref{fig:HMS-HMS-example-grid}, respectively. Although at the same time, the $q=0.60$ models also show a decrease in maximum $P_\mathrm{ZAMS}$ in Figure \ref{fig:HMS-HMS-example-grid}, because they exit the light blue squares region in Figure \ref{fig:MBH_grid_slice} due to their $\MBH/\Mdonor$ ratio. This means that the maximum $P_\mathrm{ZAMS}$ is reached around $q_\mathrm{ZAMS}\sim0.75{-}0.85$. Thus, the SMT-to-BBH-merger region in the \COHMSRLO grids determines the maximum $P_\mathrm{ZAMS}$ producing BBH mergers.

The combination of maximum $P_\mathrm{ZAMS}$ and minimum $P_\mathrm{ZAMS}$ around the same values results in $q_\mathrm{ZAMS}=0.8$ producing BBH mergers the most efficiently. Furthermore, below $q_\mathrm{ZAMS}\sim 0.50$, the minimum $P_\mathrm{ZAMS}$ from the $L_2$ instability is larger than the maximum $P_\mathrm{ZAMS}$ required for a BBH merger, thus very limited BBH mergers are produced below $q_\mathrm{ZAMS}\simlt0.5$ (see Appendix \ref{app:general_population_@_10^-4} for the ZAMS population of BBH merger progenitors at $10^{-4}\Zsun$).

Below $M_\mathrm{1, ZAMS}\approx 30 \Msun$, the physics shaping the BBH merger region deviates from above this mass limit. First, the core fraction of these stars is smaller, and mass transfer more efficiently strips the complete hydrogen envelope from these stars, even at low metallicity. Secondly, the minimum mass to form a BH becomes important. For the $q=0.6$ grid slice in Figure \ref{fig:HMS-HMS-example-grid}, BBH mergers are only formed at the tightest orbital configurations, where the secondary can sufficiently accrete material to form a BH later. This sets an upper limit on $P_\mathrm{ZAMS}$ for these mass ratios. However, as $q$ goes to unity, the shape of the BBH merger region changes. 
The companion is now more easily able to form a BH, shifting the maximum $P_\mathrm{ZAMS}$ for BBH formation upwards from 1.1 day at $q=0.6$ to 2.5 day at $q=0.85$, for $M_\mathrm{1, ZAMS}=25\Msun$. On the other hand, the shortest stable $P_\mathrm{ZAMS}$ systems are no longer able to produce BBH mergers. Due to their full stripping and additional mass loss during the supernova, the produced STAR+BH system falls within the $\MBH/\Mdonor$ region that leads to unstable mass transfer in the \COHMSRLO grid. Since \Mdonor depends on the accretion efficiency, only longer $P_\mathrm{ZAMS}$ binaries avoid unstable mass transfer in the \COHMSRLO grid. A similar effect can be seen in Figure \ref{fig:MBH_grid_slice} for a slightly higher $M_\mathrm{1,ZAMS}$ system, where for $q=0.99$ the tightest stable $P_\mathrm{ZAMS}$ (green and blue solid square markers) fall within the unstable mass transfer regime in the \COHMSRLO grid (red diamonds).
The increase in the minimum and maximum $P_\mathrm{ZAMS}$ makes this one of the only parameter regions where Case B is the first mass transfer to lead to a BBH merger, as the overlap between the black contour and the green markers in Figure \ref{fig:HMS-HMS-example-grid} for q=0.85 and q=0.99 shows. The majority of interaction occurs on the main-sequence.

\subsubsection{Linking the STAR+STAR and STAR+BH phases}

From investigating the \HMSHMS grid in this section and the \COHMSRLO grids in Section \ref{subsec:STAR_BH_phase}, we are able to draw the following conclusions regarding the formation of BBH mergers through the SMT channel: 

\begin{itemize}
    \item merging BBH progenitors come from short-period ZAMS binaries. These systems have their first interaction on the main-sequence during both the STAR+STAR and STAR+BH phases.
    \item The stability of main-sequence mass transfer in the \HMSHMS grid sets a lower period ZAMS boundary for BBH merger formation.
    \item Orbital shrinkage during the \COHMSRLO grid set an upper $P_\mathrm{ZAMS}$ boundary for the formation of BBH mergers.
    \item The accretion efficiency during the STAR+STAR phase imposes additional lower $P_\mathrm{ZAMS}$ limits, thereby restricting the formation of mergers from the shortest stable period systems depening on $M_\mathrm{1, ZAMS}$ and $q_\mathrm{ZAMS}$.
    \item The combination of minimum and maximum of $P_\mathrm{ZAMS}$ as a function of $q_\mathrm{ZAMS}$ leads to limited BBH merger formation below $q_\mathrm{ZAMS}<0.5$. 
    \item We find a limited contribution to merging BBH formation from systems where post-main-sequence, Case B, mass transfer is the first interaction in the system in both the STAR+STAR and STAR+BH phase.
\end{itemize}

\section{Why does post-main-sequence mass transfer not produce BBH mergers?} \label{sec:case_B}

\begin{figure}
    \centering
    \includegraphics[width=\linewidth]{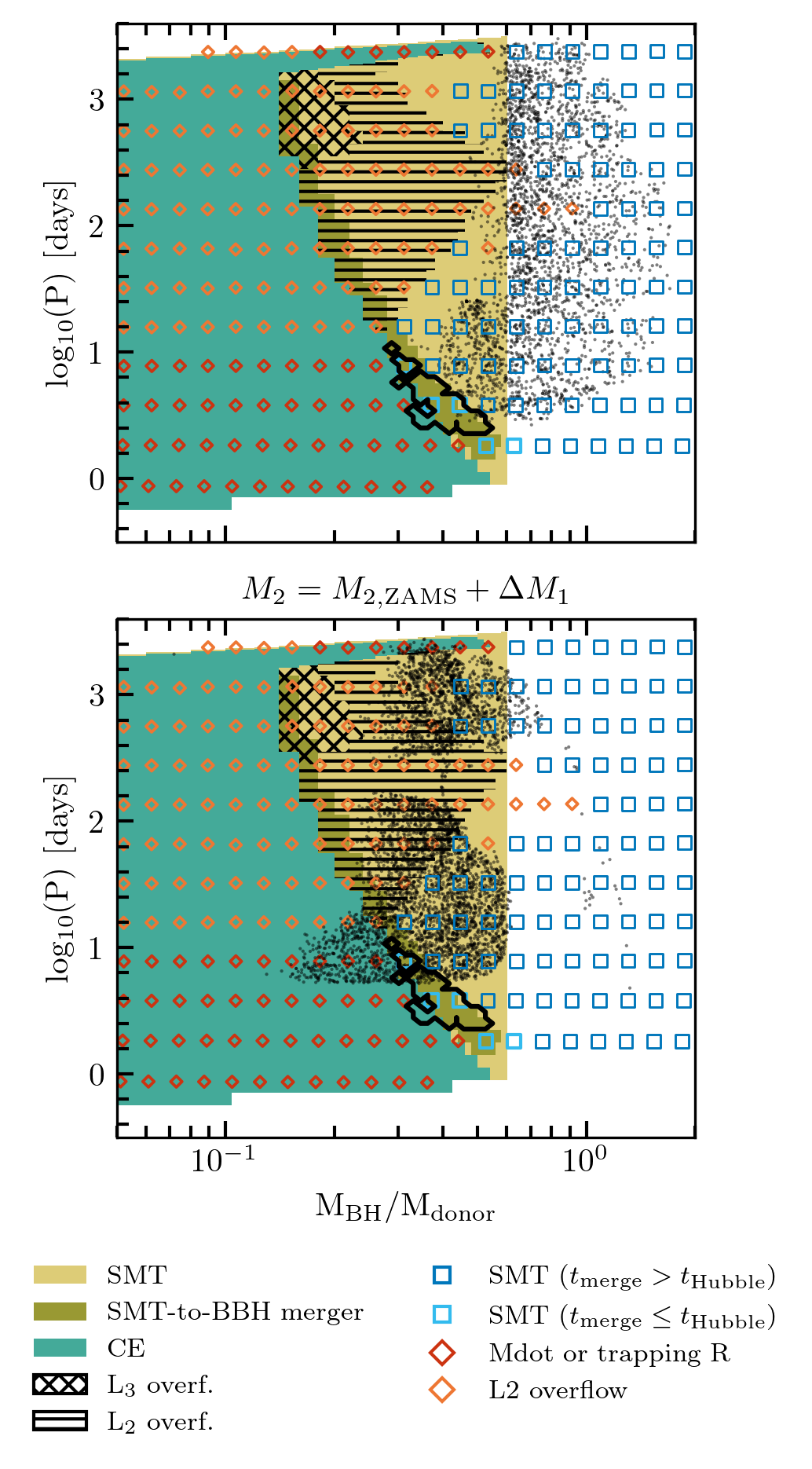}
    \caption{Both panels: \citet{Marchant+21} $\Mdonor=30\Msun$ grid with the \posydon grid at $\Mdonor\approx 33\Msun$ at $0.1\Zsun$ overlayed. Generally, the grids overlap well, although \posydon marks $L_2$ overflow as unstable.\\
    Top panel: The \posydon STAR+BH population with $\Mdonor=33\Msun\pm2\Msun$ that has undergone Case B mass transfer in the STAR+STAR phase is overlayed. There are no systems in the SMT merger region from \citet{Marchant+21} that do not fall within the SMT region of \posydon.\\
    Bottom panel: An artificial STAR+BH population where all the mass lost by the primary during the STAR+STAR phase is added to the ZAMS mass of the secondary and the orbit is adjusted analytically assuming fully conservative mass transfer and no wind loss. This approximate approach shifts lower mass companions into the $30\Msun\pm2\Msun$ regime with wider initial orbits.}
    \label{fig:marchant_comparison}
\end{figure}

The SMT channel presented in this work differs from that presented in the works using rapid population synthesis codes \citep{Neijssel+19, vanSon+22, Iorio+23, Olejak+24, Dorozsmai+24}, as well as those relying on rapid population synthesis codes for the STAR+STAR phase \citep{Bavera+21, Gallegos-Garcia+21}. As shown in Section \ref{sec:BBH_merger_formation}, the detailed binary models show that stable Case A mass transfer during both the STAR+STAR and STAR+BH phases produces nearly all ($\gtrsim95\%$) BBH mergers in the SMT channel. This is in stark contrast to rapid population synthesis codes based on \texttt{BSE} \citep{Hurley+02}, where Case B mass transfer during both the STAR+STAR and STAR+BH phases is the dominant mechanism of BBH mergers \citep{Neijssel+19, Belczynski+20a, vanSon+25}. For example, in \texttt{COSMIC} simulations, Case B mass transfer accounts for $\gtrsim 90\%$ of BBH mergers during the STAR+STAR or STAR+BH phases, while Case A during either phase only contributes $\lesssim 5\%$ to the population of BBH mergers (M. Zevin, private communication). Although the exact percentages will depend on metallicity and simulation set up, the differences in first interaction type of BBH mergers between rapid population synthesis codes and this work is substantial and presents a drastic shift in the SMT channel from Case B to Case A mass transfer.

\posydon provides a more robust treatment of Case A evolution compared to \texttt{BSE}-based rapid population synthesis codes. Case A mass transfer in \texttt{BSE}-based codes is poorly modeled due to the absence of a well-defined core and artificially produces low mass stars which do not form merging BBHs \citep[see for example, figure 4 in][]{Romero-Shaw+23}. In the \posydon detailed binary models, Case A systems detached self-consistently after mass ratio reversal or continue SMT into a Case B. Often, the Case A mass transfer is highly efficient due to strong tides in their tight orbits, although a gradient in efficiencies is present \citep[for similar effects, see][]{Sen+22}. Both the STAR+STAR and STAR+BH binary interactions in rapid population synthesis codes suffer from poor Case A mass transfer modeling, which misses the Case A contribution to the BBH merger rate in the SMT channel, though efforts are underway to address this \citep{Brcek+25}.

\subsection{Limiting Case B mass transfer in \posydon}

In this Section, we identify two important components in suppressing the formation of Case B BBH mergers in \posydon. First, the STAR+STAR Case B mass transfer sets a fundamental limit on the minimum $\MBH/\Mdonor$ ratio reachable, which is determined by how much mass can be removed from the donor star and how much of it can be accreted by the stellar companion. The former depends on the choice in core-overshooting and semi-convection, which sets the helium core size and degree of envelope stripping, while the latter is determined by the accretion efficiency.

While these parameters determine a minimum $\MBH/\Mdonor$ ratio that can be reached in the STAR+BH phase, the instability of $L_2$ outflow in the \COHMSRLO{} grid is the second component which limits the contribution of SMT Case B mass transfer to the BBH merger rate. Specifically, \posydon uses an inefficient semi-convection parameter of $\alpha_\mathrm{sc}=0.1$, as calibrated in \citet{Choi+16}, which results in a large radial expansion post core hydrogen burning. Because the star only needs to extend up to ${\sim}1.8$ times beyond their Roche lobe to reach the $L_2$ outflow condition, although in some cases less extension is required \citep[see, figure 3 in ][]{Misra+20}, most post-main sequence interactions reach this instability condition\footnote{Accretion onto the companion during the STAR+STAR phase can affect the radial evolution of the accretor on its main-sequence, see Section \ref{sec:discussion} for more details.}.

The $L_2$ outflow condition is not, on its own, limiting the formation of Case B BBH mergers. In the top panel of Figure \ref{fig:marchant_comparison}, we have overplotted the \COHMSRLO{} grid and our Case B STAR+BH population on top of the grid from \citet{Marchant+21}, where $L_2$ outflow is treated as stable, similar to grids from \citet{Gallegos-Garcia+21} and \citet{Klencki+26}. 
The region leading to BBH mergers within the Hubble time in the grid from \citet{Marchant+21} extends into the $L_2$ instability from \posydon. However, this region is not populated by STAR+BH systems. As such, to have Case B BBH mergers, both $L_2$ outflow has to be considered stable, and the minimum $\MBH/\Mdonor$ ratio has to decrease to shift STAR+BH populations into the SMT-to-BBH-merger regime with $L_2$ outflow from \citet{Marchant+21}.
The latter is set by our \HMSHMS grid; therefore, even if $L_2$ outflow were assumed to be stable, these binaries would still not occupy that region due to the mass ratio constraint imposed by the STAR+STAR phase.

\subsection{Conservative accretion experiment}
As discussed in Section \ref{subsec:STAR_BH_phase}, the minimum $\MBH/\Mdonor$ ratio for $P>10$ days is set by STAR+STAR Case B interactions. To achieve lower ratios, either $\Mdonor$ has to increase or $\MBH$ has to decrease, both of which occur in Case A systems. However, for Case B mass transfer, the helium core size is already set when reaching the end of the main-sequence, which at low metallicities determines the final formed \MBH. Thus, \Mdonor needs to increase through mass accretion, but due to rotational-limited accretion during Case B mass transfer, the mass transfer efficiency is low in \posydon. To self-consistently increase the accretion efficiency based on a physically motivated conditions during the STAR+STAR phase, new detailed binary-star model grids will need to be developed, which is beyond the scope of this work \citep[see][for development in this direction]{Xing+26}. We can, however, perform a simple experiment in which we artificially increase \Mdonor based on the total mass lost by the primary during the \HMSHMS evolution, and adjust the orbits accordingly. This drastically overestimates the final \Mdonor mass, because it assumes full accretion of all mass lost by the primary through stellar winds and mass transfer, and ignores all stellar wind mass loss from the secondary. This experiment therefore sets the high upper limit on \MBH/\Mdonor that can be reached in the most optimistic accretion scenario.

The bottom panel in Figure \ref{fig:marchant_comparison} shows the Case B population entering the \COHMSRLO{} grid with the conservative accretion assumption, implemented in the approximate way described above. As expected, lower \MBH/\Mdonor ratios are reached by $M_2$ gaining mass, but only down to $\MBH/\Mdonor=0.25$, despite our extreme accretion efficiency. This still leaves most of the post-main-sequence SMT-to-BBH-merger regime from \citet{Marchant+21}, marked in light blue, unpopulated. Even given extreme assumptions on the $M_2$ mass gain, the majority of events would still undergo Case A mass transfer during the CO-HMS phase. This demonstrates, although crudely, that under both highly inefficient and efficient mass transfer in the STAR+STAR phase, the majority of BBH mergers are not formed through Case B mass transfer\footnote{The region of SMT-to-BBH-merger in STAR+BH phase can shift depending on the angular momentum lost with the non-accreted mass leaving the binary system \citep{Klencki+26}.}.

\begin{figure}
    \centering
    \includegraphics[width=\linewidth]{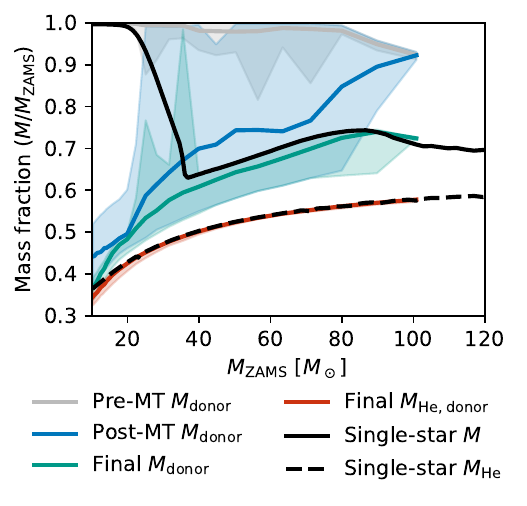}
    \caption{Fraction of the donor star remaining at different evolutionary stages in the \HMSHMS grid for Case B mass transfer at $10^{-4}\Zsun$. The mass fraction of the total star before mass transfer (gray), after mass transfer (blue), and at core-carbon depletion (green) are shown for the binary models. The figure also includes the helium core fraction of the binary models at core-carbon depletion (red). The colored regions mark the possible ranges of the fractions for different initial periods and mass ratios of binaries undergoing stable Case B mass transfer. Additionally, we show the total mass (solid black) and helium core (dashed black) fraction at carbon-depletion of the single star models.}
    \label{fig:core_1e-04_Zsun}
\end{figure}

\subsection{Role of core fraction and partial stripping}

The \MBH/\Mdonor ratio remains above 0.25 in Figure \ref{fig:marchant_comparison} due to the amount of material involved in the mass transfer. At low metallicities, a combination of the core size and partial stripping sets the maximum fraction of ZAMS mass that the donor loses through mass transfer. Figure \ref{fig:core_1e-04_Zsun} shows the helium core (red line) as a fraction of the ZAMS primary mass for all stable Case B mass transfer systems. The shading indicates the spread from different initial mass ratios and periods, which, for the final He core mass at carbon depletion, is practically negligible. Furthermore, the helium core fraction increases from 0.35 to 0.55 from $M_\mathrm{1,ZAMS}=15\Msun$ to $M_\mathrm{1,ZAMS}=100\Msun$, and follows the single-star helium core fractions (dashed line) closely. These values are mass and metallicity-dependent, and are significantly higher than $f_\mathrm{core}$ in rapid population synthesis codes \citep[e.g.][]{vanSon+22}. In detailed stellar structure models, the core size is strongly tied to the overshooting parameters, for which \posydon uses an exponential core-overshooting parameter, calibrated against the step-overshooting parameters from \citet{Brott+11} (also see the discussion in Section \ref{sec:discussion}).

Rapid population codes additionally assume the complete stripping of the hydrogen envelope during a stable Case B mass transfer, while detailed binary evolution models show that most massive primaries in stable Case B mass transfer do not fully strip their envelope during the mass transfer at low metallicity \citep{Eldridge+13, Gotberg+17, Klencki+20}. Partial stripping further reduces the available mass for the companion to gain, thus restricting the \MBH/\Mdonor ratio that can be reached. \posydon{} captures the partial stripping, as shown with the blue shaded region in Figure \ref{fig:core_1e-04_Zsun}. At the highest $M_\mathrm{ZAMS}$, the stripping is the weakest, with less than 40\% of the ZAMS mass being involved in the mass transfer in the most extreme stripping scenario. It can be as low as a few percent depending on the period and mass ratio. Towards lower ZAMS masses, the stripping becomes stronger and more of the ZAMS mass is involved in the mass transfer, up to 60\% for $M_\mathrm{1,ZAMS}=15\Msun$. For $M_\mathrm{1, ZAMS}\simlt20\Msun$, the Case B mass transfer nearly fully strips the donor star and, thus, most of the hydrogen envelope is involved in the mass transfer.

The exact amount of partial stripping depends on the primary mass and choice of semi-convection parameter \citep{Klencki+20}. \posydon uses a relatively inefficient semi-convection parameter ($\alpha_\mathrm{sc}=0.1$) which leads to efficient stripping of the envelope. Despite this, a large fraction of envelope mass remains after a Case B mass transfer, especially in the $M>25\Msun$ regime. Moreover, the core fraction sets a fundamental limit on the envelope mass that is available in the first place. Together, these effects restrict low \MBH/\Mdonor ratios from forming during STAR+STAR Case B mass transfer.

In summary, even if the Case B mass transfer efficiency during the STAR+STAR phase is increased and $L_2$ outflow during the STAR+BH phase is treated as stable, the overall population will still be dominated by systems that undergo case Case A mass transfer during the STAR+BH phase. A detailed analysis with a STAR+STAR grid with conservative mass transfer is required to understand the full effects of the accretion efficiency on limiting the Case B contribution to the BBH merger population. Such an investigation needs to be done concurrently with the treatment of the stability of $L_2$ outflow during the STAR+BH phase. Additional constraints from stellar populations can be used to better understand the earlier evolutionary phases to BBH formation and provide quantifiable predictions \citep[for example, see][at lower stellar masses]{Lechien+25, Xing+26}.

\section{Stable mass transfer across metallicity} \label{sec:metallicity_dependence}

\subsection{Origin of the metallicity dependence}  \label{subsec:origin_Z}
 
\begin{figure*}
    \centering
    \includegraphics[width=\linewidth]{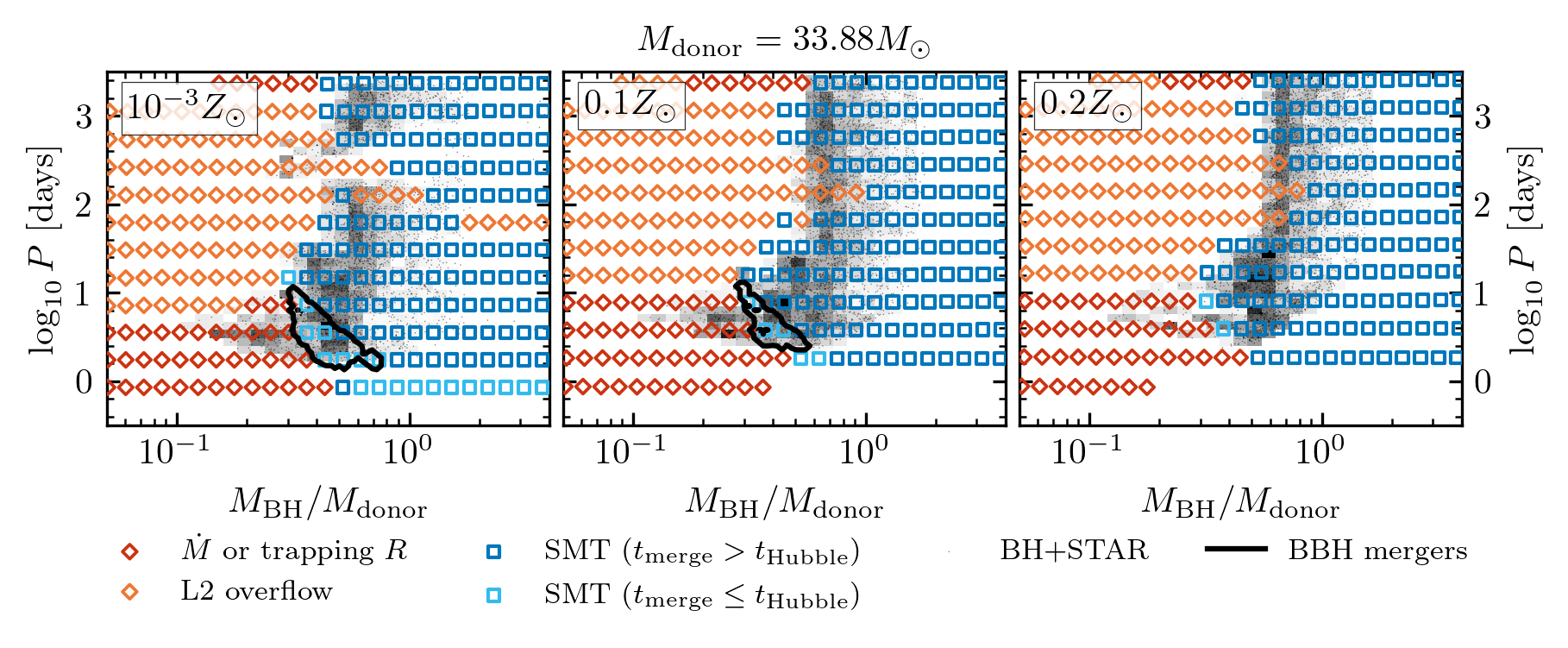}
    \caption{Grid slices of the \COHMSRLO{} grid at $\Mdonor=33.88\Msun$ for $Z=10^{-3}\Zsun$, $0.1\Zsun$ and $0.2\Zsun$. This donor mass produces BBH mergers at the highest metallicity. Above $0.2\Zsun$ the SMT channel does not produce BBH mergers in our models without natal kicks. The gray shading is the density of STAR+BH systems with $\Delta\Mdonor \pm 2\Msun$ around \Mdonor, while all systems leading to BBH mergers are outlined by the black contour.}
    \label{fig:CO-HMS_RLO_metallicity}
\end{figure*}

In the previous section, we described the formation of BBH mergers in the \posydon simulations at $10^{-4}\Zsun$. However, the majority of star formation across cosmic time does not occur at $Z=10^{-4}\Zsun$, but at higher metallicities.
Figure \ref{fig:CO-HMS_RLO_metallicity} shows the \COHMSRLO grid for $Z=10^{-3}\Zsun$, 0.1\Zsun, and 0.2\Zsun for $\Mdonor=33.9\Msun$, which is the same \Mdonor as the middle panel in Figure \ref{fig:example_Mdonor}. This donor mass is one of the few masses at which the SMT channel can still produce a BBH merger at $0.2\Zsun$ (See Section \ref{subsec:properties}).

As metallicity increases from $10^{-3}$ to $0.2\Zsun$, the parameter space in which the grid models can lead to a BBH merger within the Hubble time shrinks, and completely disappears above $0.2\Zsun$. This is the main reason for the SMT channel being restricted to $Z\leq 0.2\Zsun$, and is a result of increased orbital widening from stellar winds as metallicity increases.
The metallicity dependence affects the final orbital periods and final masses of the BHs. The increased stellar winds at higher metallicities leads to additional orbital widening during the \COHMSRLO grid and prevent the remaining star during this phase from forming of BH due to the increased mass loss.
At $\Mdonor=33.88\Msun$, the final BH masses remain sufficiently massive to enable mergers at $0.1\Zsun$ and $0.2\Zsun$, though the parameter space decreases drastically. 
The parameter space of the binaries entering the \COHMSRLO grid does not change significantly with metallicity. The only change is the minimum STAR+BH period, which shifts towards longer periods with increasing metallicity. This is a consequence of additional orbital widening before the first supernova due to stellar winds and increased stellar radii at ZAMS, preventing stable interactions at the tightest orbital configurations. As a result, the short-period SMT-to-BBH-merger models at $P\sim1.8$ days and $0.1\Zsun$, marked in light blue in the middle panel of Figure \ref{fig:CO-HMS_RLO_metallicity}, are not populated by STAR+BH systems (grayscale and dots). However, this effect is minor compared to the shrinkage of parameter space in the \COHMSRLO grid due to stellar wind mass loss.

At donor masses more massive than $\Mdonor=33.9\Msun$, stellar winds have a stronger effect and cause more orbital widening before and after the first supernova. As a result, the SMT-to-BBH-merger region in the \COHMSRLO grid already disappears at $0.1\Zsun$, preventing BBH mergers from forming through SMT at higher metallicities.
A similar effect occurs at donor masses below $33.9\Msun$, where the SMT-to-BBH-merger region in the \COHMSRLO grid is eliminated with increasing metallicity. Although these lower-mass stars experience weaker stellar winds, the increased mass loss and orbital widening compared to lower metallicities, nonetheless, lead to the formation of less massive and wider BBH systems. Gravitational wave emission is insufficient to merge these systems within the Hubble time. 
This implies that the BBH mergers from the SMT channel have a minimum and maximum BH mass that is determined by the stellar wind mass loss in the \COHMSRLO grid (see Section \ref{subsec:properties}), although this can be further restricted by the supernova remnant mass prescription linking the final core masses to BH masses (See Appendix \ref{app:other_SN_prescriptions} for other prescriptions).

To summarize, we find that orbital widening due to stellar winds during the STAR+BH phase, modeled in the \COHMSRLO grid, leads to a metallicity dependence in the BBH mergers through SMT, where no mergers occur above $0.2\Zsun$ without a natal kick (see Section \ref{sec:natal_kicks} for the population with kicks). We find additional sub-dominant effects from larger stellar radii at ZAMS \citep[see also][]{vanSon+25} and decreased mass transfer stability during the STAR+STAR phase, as metallicity increases. Orbital widening due to stellar wind mass and angular momentum loss is the dominant metallicity effect on BBH mergers in the SMT channel.

\subsection{Population Properties} \label{subsec:properties}

\begin{figure*}
    \centering
    \includegraphics[width=\linewidth]{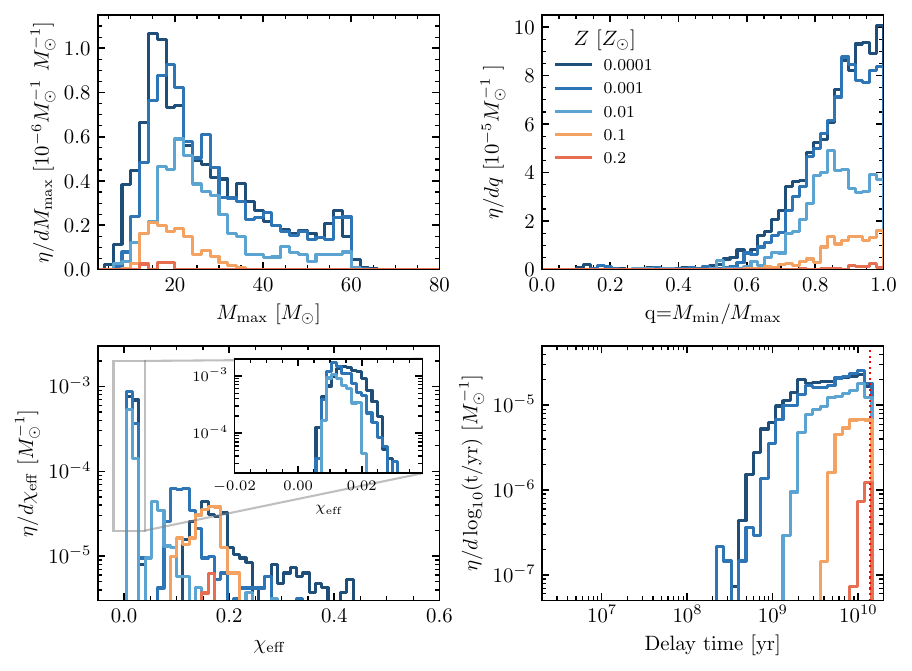}
    \caption{Properties of BBH mergers with a delay time less than the Hubble time per metallicity without convolution with a cosmic star formation history. Top left: the mass distribution of the more massive BH in the merger. Top right: The mass ratio distribution. Bottom left: \chieff distribution. Bottom right: the delay time distribution of the population. The red dotted line indicates the age of the Universe.}
    \label{fig:merger_pop}
\end{figure*}

\subsubsection{Mass Distributions}
The effect of metallicity is strongly visible in the metallicity-specific BBH merger populations in Figure \ref{fig:merger_pop}. 
The SMT channel only produces BBH mergers at $Z\leq0.2\Zsun$, when no natal kick is applied. Above this metallicity, the SMT channel cannot produce BBH mergers due to the parameter space for a merger within the Hubble time in the \COHMSRLO grid closing. At the lowest metallicities ($Z<0.01\Zsun$), we find that the SMT channel covers the full range of BH masses from 3\Msun to 65\Msun (top-left panel), with a main peak around ${\sim}15\Msun$ and a secondary peak at ${\sim}60\Msun$ from pulsational-pair instability. In the primary BH mass distribution, three effects are visible as metallicity increases. First, the strongest effect is the disappearance of high-mass BBH mergers at higher metallicities. Although massive BHs are still able to form and interact through SMT, their orbits are too wide to merge within the Hubble time. This shifts the maximum BH mass down to ${\sim}20\Msun$ for $Z=0.2\Zsun$.
The second effect is the decrease in efficiency for BBH merger formation, which drops an order of magnitude from ${\sim}10^{-5}\Msun^{-1}$ to ${\sim}10^{-7}\Msun^{-1}$ between $10^{-4}\Zsun$ and $0.2\Zsun$.
The final effect is the increase of the lowest $\Mmax$ from $3\Msun$ to $10\Msun$, as metallicity increases, since stellar winds also start to affect the merger time for lower mass stars.
Without a natal kick, the SMT channel does not efficiently produce BBH mergers below ${\sim}15\Msun$, which implies that a SMT population, without a natal kick, will be unable to produce the observed $10\Msun$ peak in gravitational wave data \citep{Abac+25a}. We discuss the effect of natal kicks in Section \ref{sec:natal_kicks}, but the population presented here is minimally affected by a natal kick prescription that uses fallback or a BH mass scaling.

\subsubsection{Mass Ratio}
Independent of metallicity, the mass ratio of the SMT channel with Eddington-limited accretion has a preference for more equal mass ratios. As discussed in Section \ref{subsec:STAR+STAR}, the initial mass ratios are limited to $q_\mathrm{ZAMS}\simgt0.5$ due to Case A mass transfer stability in the STAR+STAR phase. In Appendix \ref{app:general_population_@_10^-4}, we discuss the populations at each evolutionary phase at $10^{-4}\Zsun$, and find that mass ratio reversal causes a Gaussian-like peak around $M_\mathrm{2,BH}/M_\mathrm{1, BH}=1$ (see orange line in bottom left panel in Figure \ref{fig:1e-4_evolution_properties}). If instead shown as $\Mmin/\Mmax$, without knowledge of which BH formed from the initially more massive star, we find a mass ratio distribution that peaks at 1, nearly independent of the metallicity, as shown in Figure \ref{fig:merger_pop}. As discussed in Section \ref{sec:metallicity_dependence}, the parameter space for SMT-to-BBH-mergers shrinks with increasing metallicity, only leaving systems with equal masses to merge within the Hubble time due to more efficient gravitational wave emission.

\subsubsection{$\chieff$ distribution}
The $\chi_\mathrm{eff}$ distributions in the bottom left panel of Figure \ref{fig:merger_pop} show two distinct peaks in the SMT channel, one at $\chieff=0$ and another at $\chieff\approx0.1{-}0.15$. Below $Z<0.01\Zsun$, there is an additional long tail towards high $\chieff$, which is a result of tight-orbit high-mass systems, which are rapidly rotating due to strong tides \citep{Marchant+16}. Their tight initial orbits cause strong rotation and its associated mixing, keeping the star nearly chemically homogeneous, avoiding the red-supergiant branch. It retains a high angular momentum reservoir until collapse, producing BHs with relatively high \chieff. In our study, these binaries exhibit mass transfer without triggering instability and, therefore, are placed into the classification of SMT, though their rate is 2 orders of magnitude lower than the main peaks at the same metallicity.

The main $\chieff=0$ peak is caused by long-period systems where the progenitor has a slower initial rotation. This peak only occurs for lower metallicity systems ($Z<0.01\Zsun$), because at higher metallicity stellar winds widen these long-period systems, preventing them from merging within the Hubble time. As a result, only the $\chieff \sim0.15$ peak remains at higher metallicities, $Z\simgt0.01\Zsun$. These are very close systems at ZAMS that are still able to merge, where both stars are spun up due to tidal interactions in the HMS-HMS grid. The primary and secondary retain sufficient angular momentum until they collapse, forming BHs with individual spins around ${\sim}0.15$. Thus, this produces a metallicity dependence in the $\chieff$ distribution, where it increases with increasing metallicity.

\subsubsection{Delay Time Distribution}
Finally, in the bottom-right panel in Figure \ref{fig:merger_pop}, we show the delay time distribution. The SMT channel produces long delay times, that increase even further with metallicity due to orbital widening from increasing stellar winds. However, already at $Z=10^{-4}\Zsun$, the minimum delay time is 0.2 Gyr, drastically impacting the contribution of the SMT channel to the observed BBH merger population. These findings are in line with \citet{Klencki+26} who find a minimum delay time of ${\sim}1$ Gyr at $0.1\Zsun$, originating from a post-main sequence donor system. At the same metallicity, we find ${\sim}4$ Gyr due to only having main-sequence donors, which results in a wider final separation at BBH formation compared to more evolved donors \citep[see][for more details]{Klencki+26}. These long delay times imply an intrinsic BBH merger rate density from the SMT channel that is increasing to the present day ($z=0$), which we will discuss in a follow-up work.

\section{The effect of natal kicks} \label{sec:natal_kicks}

\begin{figure*}
    \centering
    \includegraphics[width=\linewidth]{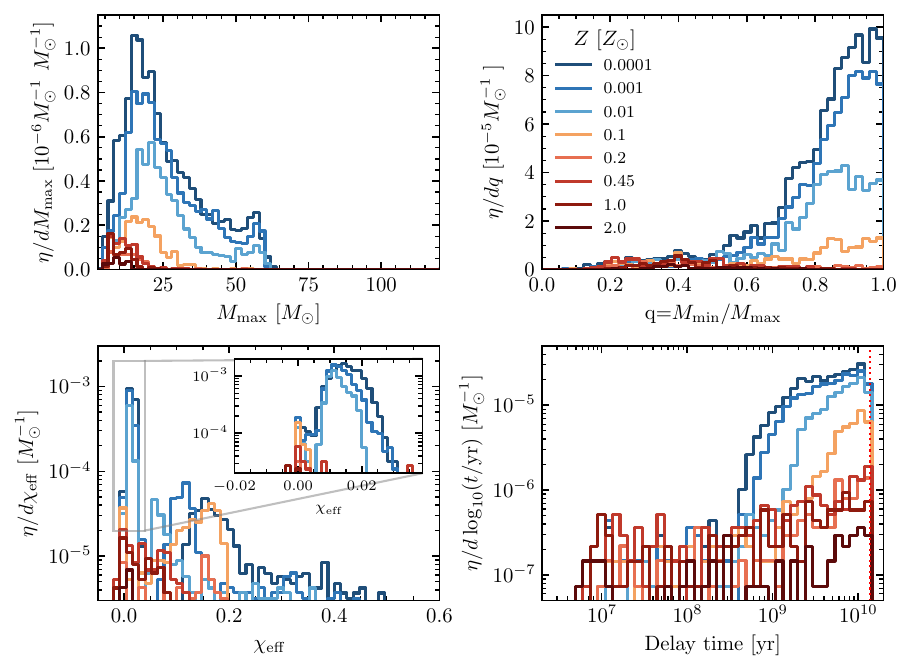}
    \caption{BBH mergers populations with a mass-scaled natal kick. The panels are the same as Figure \ref{fig:merger_pop}. An additional population of low mass BHs with $q\sim0.2{-}0.6$ and slightly positive \chieff appears, which allows for the formation of BBH mergers at $Z>0.2\Zsun$.}
    \label{fig:merger_obs_kick}
\end{figure*}

During a supernova, asymmetry in the ejecta can induce a natal kick on a neutron star \citep{Janka+94, Burrows+95, vandenHeuvel+97}, which several studies have tried to constrain the imparted velocity \citep[e.g.][]{Hobbs+05, Disberg+25}. A similar effect occurs during the formation of a BH if material is ejected asymmetrically, though the velocity of this kick is uncertain \citep[e.g.][]{Janka+24, Burrows+25}. In this section, we show the effects that a natal kick prescription has on the formation of BBH mergers from the SMT channel.

A fallback-modulated or a mass-scaled natal kick are commonly used prescriptions in BBH population synthesis \citep[e.g.][]{Fragos+10, Zevin+17, Belczynski+16, Mapelli+18, Eldridge+19a, Broekgaarden+22a, Iorio+23, Olejak+24}, which predominantly affect the low-mass end of the BH mass spectrum. The natal kicks are drawn from a Maxwellian distribution with $\sigma=265$ km/s following \citet{Hobbs+05} \citep[although, see the correction by][]{Disberg+25}, which receives a correction based on the fallback or a reduction by $1.4/\MBH$. For the populations in Section \ref{subsec:properties}, most merging BBHs form in the mass regime for direct collapse. Thus, only a negligible natal kick can be imparted on the BHs though these two kick prescriptions, and our zero natal kick assumption holds in the high-mass regime. A small \citet{Blaauw+61} kick is, nonetheless, imparted by the assumed instantaneous mass loss of $0.5\Msun$ through neutrino emission during the BH formation at all masses. 

Figure \ref{fig:merger_obs_kick} shows the merger populations with a mass-scaled \citet{Hobbs+05} natal kick, predicted by \posydon. We apply a natal kick prescription during both supernovae. The kick imparted during the second supernova can significantly affect the delay time of the system.
The introduction of eccentricity from the natal kick at the formation of the second-born BH allows binaries that were too wide before the supernova to merge within the Hubble time, specifically lower mass BBH systems and at $Z>0.2\Zsun$. These newly merging systems have received a kick in the opposite direction to the orbital velocity. Applying a kick prescription does not drastically alter the population already present without the natal kick, as expected due to their masses.
The main effect on the BBH merger population is that the natal kick introduces an additional peak at $\Mmax\sim5{-}10\Msun$ with mass ratios around $\Mmin/\Mmax\sim0.2{-}0.6$, which aligns more closely with observations of the low-mass peak containing more unequal mass ratio mergers \citep{Abac+25a}. However, the exact location of the mass and mass ratio peak per metallicity is dependent on the assumed remnant mass prescription (see Appendix \ref{app:other_SN_prescriptions}).

Additionally, the bottom left panel in Figure \ref{fig:merger_obs_kick} shows that the \chieff distribution of the new high-metallicity kicked population is slightly asymmetric around 0 with a preference skew towards positive \chieff values, for which some observational evidence might be seen \citep{Abac+25a}.
The kicked population, generally, merges more rapidly than the population without kicks. These progenitors of these BBH mergers are not constrained to Case A mass transfer being their first interaction. In the bottom right panel in Figure \ref{fig:merger_obs_kick}, we see that the population with long delay times, which comes from BBHs with no kicks, remains present, as expected. These systems contain massive BHs, which receive weak natal kicks in the mass-scaled kicked population. Although the features in these single-metallicity populations show similarities to the observed BBH population, a more thorough investigation of these features in a cosmic population context is required, as convolution with the star formation history impacts the rate density and mass distribution.

\section{Discussion} \label{sec:discussion}

The dominance of Case A mass transfer in the SMT channel introduces several dependencies on stellar and binary evolution physics. The most significant uncertainties stem from semi-convection, stellar wind mass loss, and model assumptions of the accretor after the first SN. For example, the semi-convection parameter choice determines several important aspects in the interaction, such as the stability of Case A mass transfer and the radial expansion post-main-sequence, which indirectly determines the efficiency of stripping during mass transfer. Secondary uncertainties arise from core-overshooting and the assumed initial period distribution.  In particular, we focus on the evolutionary appearance of BBH progenitors, the interplay between stability of mass transfer and single-star physics assumption, and how the \posydon modeling assumptions might affect the presented results.

\subsection{Semi-convection Parameter}
The semi-convection parameter plays a role during two moments in the evolution to a BBH merger through SMT. First, it determines the radial response of the accretor during the STAR+STAR phase. And secondly, it determines the post-main-sequence radial expansion, which affects the amount of stripping during the STAR+STAR phase and the stability of mass transfer in the STAR+BH phase. Following the calibration of the MIST2 models (Dotter et al., in prep), \posydon uses a relatively inefficient semi-convection choice of $\alpha_\mathrm{sc}=0.1$ \citep{Andrews+25}. In contrast, observations of single stars in the Small Magellanic Cloud suggest that semi-convection needs to be efficient ($\alpha_\mathrm{sc}\simgt 10$) to reproduce the number of blue supergiants \citep{Schootemeijer+19}. This has implications for the accretor response during mass transfer and the radial expansion of the donor, which we discuss in the next two paragraphs.

During the STAR+STAR phase, the stability of mass transfer is primarily determined by the $L_2$ outflow condition during a contact phase, as discussed in Section \ref{subsec:STAR+STAR}, although a contact phase does not always lead to the $L_2$ instability, especially when the companion can regain thermal equilibrium, which is easier at high main-sequence companion masses \citep{Henneco+23}. The contact phases that do reach the $L_2$ outflow condition originate from the accretor expanding as a response to incoming material from the donor star, which is closely related to the semi-convection efficiency \citep{Wellstein+01, Menon+21}. The inefficient semi-convection in \posydon limits the rejuvenation of a late-stage main-sequence companion \citep{Braun+95}, causing it to expand more easily as a response to accreted material. More efficiency semi-convection would induce rejuvenation, keeping the accretor compact until the primary reached core-collapse, reducing its expansion and possibly permitting stable interaction in systems with shorter initial periods \citep{Wellstein+01}. This would extend the region of $P_\mathrm{ZAMS}$ leading to BBH merger through SMT. Thus, the \posydon binary grids likely represent a conservative scenario for the survivability of main-sequence mass transfer due to the inefficient semi-convection parameter. A more efficient semi-convection parameter could lead to more Case A STAR+STAR systems reaching the STAR+BH phase, potentially increasing the number of BBH merger through the SMT channel.

Semi-convection also determines the post-main-sequence radial expansion of the primary star \citep{Georgy+13} and, through this, the amount of (partial) stripping \citep{Klencki+20}. This is crucial for the exclusion of Case B BBH progenitors, as the amount of stripping determines the maximum mass ratio change during the STAR+STAR phase, as discussed in Section \ref{sec:case_B}. The inefficient semi-convection in \posydon should lead to relatively efficient stripping, but despite this, we still find partial envelope stripping. With more efficient semi-convection ($\alpha_\mathrm{sc}\simgt10$), the donor star is able to thermally readjust more rapidly, and partial stripping would be more common. Thus, a smaller amount of mass would be transferred to the companion, making the formation of STAR+BH systems that can merge within the Hubble time through SMT more difficult.  While the previous work on semi-convection and mass transfer stability \citep{Wellstein+01, Schootemeijer+19} considers a slightly lower mass range of massive stars at the high end of the metallicity range for BBH merger formation in \posydon, the effects remain qualitatively similar for the higher-mass systems that dominate our BBH merger population. More efficient semi-convection will make it more difficult to produce BBH mergers through Case B mass transfer, further supporting the dominance of Case A in BBH merger formation from the SMT channel.

\subsection{Core-overshooting}
An increase in the mass budget during Case B mass transfer can also be achieved by a lower core-overshooting parameter \citep[e.g.][]{Kaiser+20}, which would produce a larger envelopes, smaller core masses and, thus, also smaller BH masses \citep{Temaj+24}. \posydon uses an exponential core-overshooting prescription calibrated against the results from \citet{Brott+11}, which, in turn, is calibrated for ZAMS masses between $10\Msun$ to $20\Msun$ based on rotating stars in the LMC. The majority of BBH mergers in this work originate from ZAMS masses above $30\Msun$, where the core-overshooting is less constrained by observations. While a smaller core-overshooting in the high-mass regime would allow smaller cores to form and possibly allow for BBH mergers through Case B mass transfer during the STAR+STAR and STAR+BH phases if $L_2$ outflow is considered stable, observational evidence disfavors lower overshooting parameters \citep{Schootemeijer+19, Higgins+19, Kaiser+20}. Given the higher core-overshooting parameter in \posydon, this effect is likely secondary to the semi-convection choice but this should be investigated in more detail to determine the impact on the BBH merger population from the SMT channel.

\subsection{Stellar Wind Mass Loss}

Besides the internal single-star physics, the stellar wind mass loss plays a crucial role in the orbital widening in the \COHMSRLO grid as a function of metallicity. For hot H-rich stars, \posydon uses the wind mass loss scheme from \citet{Vink+00} with a $Z/\Zsun^{0.68}$ metallicity dependence \citep{Vink+01}. Generally, more recent updated prescriptions for hot massive star indicate weaker stellar winds and a stronger decrease in the stellar wind mass loss with metallicity than currently implemented \citep[e.g.,][]{Sabhahit+22, Sabhahit+23, Gormaz-Matamala+22, Gormaz-Matamala+24}. This implies that the parameter space for BBH merger through SMT in the \COHMSRLO grid will remain present up to higher $Z$. At the same time, STAR+BH systems entering this grid will likely have a shorter period compared to the current population. This will have a strong impact on the final BBH merger population and its metallicity dependence.

\subsection{Treatment of companion in \posydon}

Although \posydon self-consistently models the binary in each of the binary grids, the companion, after the first supernova, is matched to the track of a single star\footnote{This allows the companion to be matched to a more evolved star, not only to ZAMS.}. During this phase, the effects of stellar spin on the orbit are included through the tidal spin-orbit coupling, but effects of rotation on the stellar structure are not considered\citep[For more details, see section 8.1.2 in ][]{Fragos+23}. Furthermore, the matching to a single star track removes any possible structure changes caused by mass accretion from the later evolution of the companion. This might cause the companion to expand on the main-sequence, despite its original abnormal envelope-to-core ratio. \citet{Xu+25a} have shown that self-consistently modeling the accretor from the STAR+STAR phase allows it to remain compact on its main-sequence. These systems undergo stable Case B mass transfer during the STAR+BH phase at more extreme mass ratios and shorter periods compared to a STAR+BH system with the star initiated at ZAMS \citep{Xu+25a}. These are not the traditional Case B interactions at longer orbital periods discussed in Section \ref{sec:case_B}, but vecause \posydon replaces the accretor after the first supernova, the effects from accretion can result in stable Case A systems becoming unstable and unstable Case A systems undergoing stable Case B. However, continuing the modeling of the accretor is difficult in the context of natal kicks, where the induced eccentricity affect the moment the first mass transfer initiates. This highlights the importance of detailed modeling of the accretor in context of BBH merger formation.

\subsection{Broader consequences}

The Case A-dominated formation pathway and metallicity dependence of BBH mergers through SMT have several additional consequences beyond the merger rate and delay time distributions discussed in Section \ref{sec:metallicity_dependence}. These implications concern the observable properties of progenitor systems, the prevalence of classical Wolf-Rayet phases, and the sensitivity of predicted merger rates to assumptions about the initial binary population. We discuss each of these aspects below.

Traditionally, the donor stars in BBH merger progenitors are expected to become fully-stripped after mass transfer and appear as a classical Wolf-Rayet star \citep[e.g.][]{Belczynski+12}. A key consequence of the Case A formation and low-metallicity preference of BBH mergers through SMT is the absence of a fully stripped phase in the STAR+STAR and STAR+BH phase of the BBH merger progenitors in the non-kicked population. Due to the adjustment of the convective core,  mass ratio reversal during the main-sequence, and low metallicity preference, the donor stars generally retain a large hydrogen envelope until core-carbon depletion, as shown in the example binary in Figure \ref{fig:example_binary}.

Case B mass transfer systems can experience a Wolf-Rayet phase depending on the metallicity, but these generally do not lead to BBH mergers in our simulations with \posydon. Additionally, at the lowest metallicity considered, a large hydrogen envelope is also retained after a Case B mass transfer (See Section \ref{sec:case_B}), which prevent the presence of a classical Wolf-Rayet phase as well. Combined with the low-metallicity formation of BBH mergers for the non-kicked population, the population of BBH merger progenitors are unlikely to have undergone a classical hydrogen-free Wolf-Rayet phase, though a more detailed investigation is needed. Our population still allows Wolf-Rayet stars to form BHs and for BBH merger progenitors with a strong natal kick from higher metallicity environments to have experienced a classical Wolf-Rayet phase.

The preference for short initial periods from the Case A population introduces a dependence on the initial period distribution for its BBH merger efficiency in a single-metallicity population. Figure 7 in \citet{Bavera+21} shows the initial orbital periods sampled from a log-uniform and their extrapolated \citet{Sana+12} distribution. At $P<10$ days, the difference between these distributions is a factor 3. With new results from \citet{Sana+25} for the SMC, the period distribution might even have a slightly positive slope with $\log(P)$, introducing an even larger uncertainty in the contribution of the shortest-period systems. Thus, the choice of period distribution will impact the BBH merger efficiency per metallicity from the SMT channel.

\section{Conclusion} \label{sec:conclusion}

Using binary population synthesis in \posydon, we have explored the formation of binary black hole mergers through stable mass transfer (SMT), providing the first exploration of SMT with detailed binary models from ZAMS to merger over a full range of masses and metallicities.

We find that main-sequence, Case A mass transfer dominates the formation of BBH mergers through the SMT channel. This shows the need for detailed binary models in understanding the formation of BBH mergers, as this population is not currently captured in rapid population synthesis codes. Depending on the metallicity, these Case A BBH progenitors retain a hydrogen envelope up to core-carbon depletion despite undergoing mass transfer, avoiding a potential Wolf-Rayet phase and its associated strong mass loss. We also demonstrate that the formation mechanism of BBH mergers through SMT in rapid codes, which is predominantly Case B post-main-sequence mass transfer, does not produce a significant number of BBH mergers within the Hubble time in \posydon, unless different assumptions on mass accretion efficiency during the STAR+STAR phase, mass transfer stability, core-overshooting, and semi-convection parameters are made.

Within the SMT channel, we find a dominant sub-population that is robust against our natal kick assumptions (see Section \ref{subsec:properties}). These BBH mergers have, independent of metallicity, mass ratios distributions peaking around $q\sim0.8{-}1$. The primary BH mass distribution covers a wide range of masses at low metallicity ($3{-}65\Msun$), but shrinks to only systems at ${\sim}20\Msun$ towards $0.2\Zsun$ due to orbital widening from stellar wind mass loss during the STAR+BH phase. The increase in stellar winds also leads to metallicity-dependent delay time distributions, with no mergers within the Hubble time occurring above $Z\simgt0.2\Zsun$. Finally, we find peaks at $\chieff=0$ and $\chieff=0.15$ in the \chieff distributions per metallicity originating from long-period and short-period progenitors, respectively. As metallicity increases, long-period binaries no longer contrinbute to the BBH merger population and the $\chieff=0$ peak disappears at higher metallicities. This population of BBH mergers produces unique features that generally remain similar across supernova remnant mass prescriptions (Appendix \ref{app:other_SN_prescriptions}).

Going from a zero-kick population to one with a mass-scaled natal kick adds a subpopulation of BBH mergers. The natal kicks and their mass-scaling introduce eccentricity to lower-mass ($M\simlt20\Msun$) and higher metallicity ($Z>0.2\Zsun$) BBH binaries that were previously excluded due to their long delay time, allowing them to merge within a Hubble time, often faster than the zero-kick population. Moreover, this subpopulation has a preference for unequal mass ratios ($q\sim0.2-0.8$) and $\Mmax \sim 10\Msun$, although dependent on the choice of supernova remnant mass prescription.

This work provides a next step in understanding how detailed binary models provide self-consistent predictions for the formation of BBH mergers within the SMT channel. Furthermore, it provides further insight into the single and binary physics that are important in shaping the final population properties per metallicity. The two populations found in this work provide unique and interesting features in their metallicity-specific population properties, but further work is required to interpret them in the context of the observed population of BBH mergers.

\begin{acknowledgements}

We would like to thank Emmanouil Zapartas and Zepei Xing for the discussion and  helpful comments on the manuscript. The \posydon project is supported primarily by two sources: the Swiss National Science Foundation (PI Fragos, project number CRSII5\_213497) and the Gordon and Betty Moore Foundation (PI Kalogera, grant award GBMF8477). 
MMB was supported by the project number CRSII5\_213497, and the Swiss Government Excellence Scholarship No. 2024.0234.

\end{acknowledgements}

\bibliographystyle{aa}
\bibliography{BBH_SMT_channel}

\begin{appendix}

\section{Population properties at $10^{-4}\Zsun$} \label{app:general_population_@_10^-4}

\begin{figure*}
    \centering
    \includegraphics[width=\linewidth]{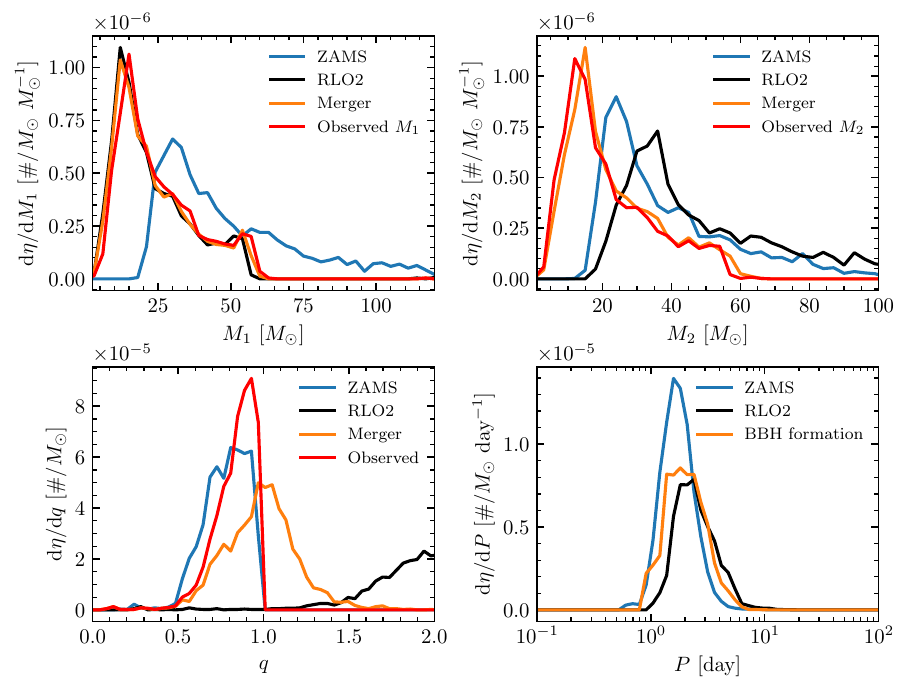}
    \caption{Properties of BBH mergers at $Z=10^{-4}\Zsun$ in the SMT channel, as probability density functions of the number of BBH mergers per $\Msun$ of star formation. $M_1$ and $M_2$ indicate the initial at ZAMS of more (top-left) and less massive (top-right) star, respectively. However, the observed $M_1$ and $M_2$ in a BBH merger are defined based on their mass at merger (red line). The mass ratio is defined as $q=M_2/M_1$, except for the observed $q$, where $q=\min(M_\mathrm{BH})/\max(M_\mathrm{BH})$} 
    \label{fig:1e-4_evolution_properties}
\end{figure*}

When discussing BBH merger formation, we distinguish explicitly between the population synthesis and the \posydon\ grids. The grids show how binaries evolve during specific evolutionary phases, independent of each other, such as STAR+STAR and STAR+BH, whereas the population synthesis links all phases together to create an astrophysical population. Both perspectives are required to understand how BBH mergers are formed through the SMT channel. Figure~\ref{fig:1e-4_evolution_properties} shows the general properties of the $10^{-4}\,\Zsun$ population across the STAR+STAR, STAR+BH, and BBH phases. 
The evolutionary outcome of the HMS-HMS grid, where the STAR+STAR interaction takes place, determines the properties of the binaries subsequently undergoing a STAR+BH mass transfer phase, as modeled in the CO-HMS\_RLO grid\footnote{The CO-HMS\_RLO grid contains binaries consisting of a star and compact object companion. Non-interacting binaries and the evolution before RLO are excluded from this grid. See \citet{Fragos+23} and \citet{Andrews+25} for more details.}. We begin our analysis with the latter evolutionary phase, as this enables us to identify the properties of binaries entering the CO-HMS\_RLO grid and quantify the effect of this grid on the formation of BBH mergers. 

\textbf{Orbital Period:} The bottom-right panel demonstrates that nearly all binaries have periods less than 10 days throughout their evolution. At ZAMS, the BBH merger progenitors span $P_\mathrm{ZAMS}{\sim}0.7{-}7$ days. The STAR+STAR binary interactions and first  supernovae widen binary systems to periods up to ${\sim}30$ days, while also increasing their minimum period to $P\gtrsim1$ day at the start of Roche lobe overflow onto the BH. We discuss the minimum period of systems entering the \COHMSRLO in Section \ref{subsec:STAR_BH_phase}.
The subsequent STAR+BH mass transfer and the second SN shrink the orbit, resulting in BBH progenitors with periods between 0.9 and 10 days. The minimum period is consistent with \citet{Klencki+26}, who found ${\sim}0.8$ days (separation of 8-10\Rsun). These periods determine the delay times of the SMT-channel BBH mergers (see Section \ref{subsec:properties}). Because the binaries remain in a tight orbital configuration throughout their evolution, nearly all BBH progenitors experience Case A mass transfer in both phases.

\textbf{Mass distributions:} At $Z=10^{-4}\Zsun$, the primary BH mass distribution spans the full range from the minimum BH mass at $2.5 \Msun$ to the onset of the pair-instability supernova regime at ${\sim}60$~\Msun with a pile-up at ${\sim}55\Msun$ due to pulsational pair-instability mass loss. While BBH mergers with masses above the PISN mass gap ($M > 100$~\Msun) are produced through the SMT channel, their rate density per unit star formation is low compared to mergers below the mass gap. The corresponding ZAMS masses of the BH progenitors range from $20~\Msun$ to $300~\Msun$, with the distribution peaking at approximately $30\,\Msun$. By definition, the ZAMS masses of secondary exhibit a lower range, spanning $15$ to $300\,\Msun$. The mass transfer in the STAR+STAR phase shifts the $M_2$ distribution upwards and produces in 51\% of the systems a larger second-born BH than the first-born BH. The mass ratio reversal means that the primary star does not produce the more massive component in the BBH merger. As such, comparing the intrinsic primary mass distribution at merger (orange lines) to the \Mmax distribution (red lines) shifts the ${\sim}10\Msun$ peak towards slightly higher values, with the opposite effect of $M_2$ and \Mmin.

\textbf{Mass ratio:} At $10^{-4}\Zsun$, 51\% of progenitors undergo mass-ratio reversal, producing a more massive second-born BH. The bottom-left panel in Figure \ref{fig:1e-4_evolution_properties} illustrates this with the observed (red) and intrinsic (orange) mass ratio distributions. 
At ZAMS, the binaries have $q\simgt0.5$, peaking at $q=0.85$. During the subsequent STAR+STAR mass transfer phase, efficient accretion onto the secondary\footnote{The efficient accretion occurs for BBH progenitors due to their short initial periods, which allows tides to spin down the accretor during the STAR+STAR mass transfer, avoiding rotation-limited accretion. Depending on the binary masses, longer periods are less efficient due to weaker tides.} inverts the mass ratio, reaching up to $M_2/M_1\approx4$ by the time $M_1$ has collapsed to form the first BH.
The majority of mass ratios remain inverted during the STAR+BH phase and second BH formation, as Eddington-limited accretion onto the first-born BH constrains its mass gain during the mass transfer.

\section{The effect of the remnant mass prescription} \label{app:other_SN_prescriptions}

In this Section, we show the effect of using two additional remnant mass prescriptions: the \citep{Patton+20} remnant mass prescription with the \texttt{N20} engine \citep{Saio+88, Nomoto+88, Shigeyama+90} and the \citet{Sukhbold+16} prescription with the \texttt{W20} engine \citep{Woosley+97}.
We show the BBH merger properties with and without the inclusion of mass-scaled natal kicks.
 
\subsection{No natal kick populations}

\begin{figure*}
    \centering
    \includegraphics[width=\linewidth]{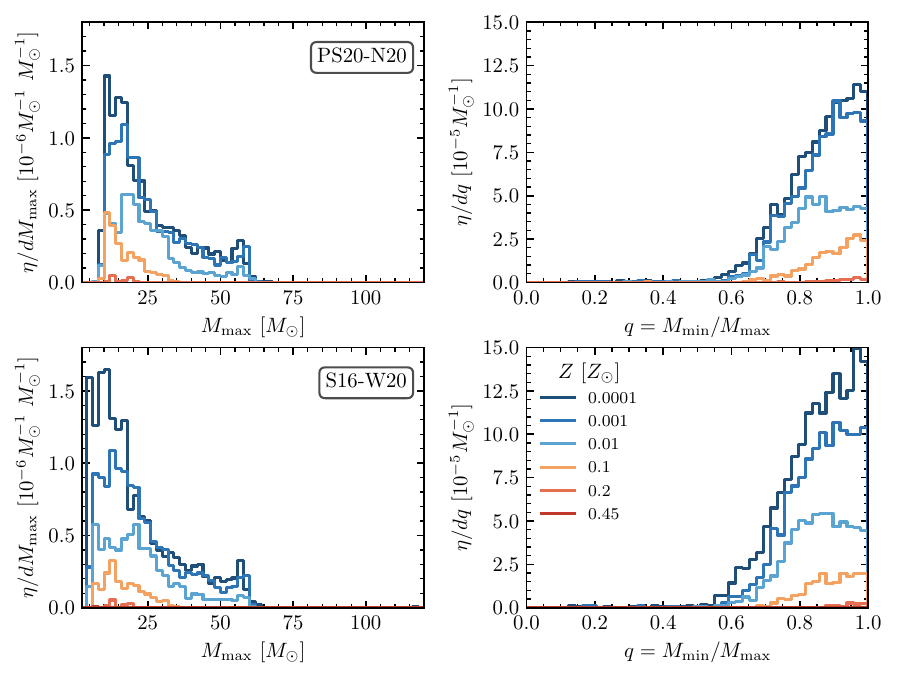}
    \caption{Mass and mass ratio distributions per metallicity for the \citet{Patton+20} remnant mass prescription with the \texttt{N20} engine \citep{Saio+88, Nomoto+88, Shigeyama+90} and the \citet{Sukhbold+16} prescription with the \texttt{W20} engine \citep{Woosley+97} on the top and bottom rows, respectively. Both prescriptions produce more BBHs compared to the \citet{Fryer+12} \texttt{delayed} prescription in Figure \ref{fig:merger_pop}. The colors indicate different metallicities, and no natal kick is applied.}
    \label{fig:other_SN_prescriptions}
\end{figure*}

A remnant mass prescription links the stellar properties at carbon depletion to the compact object properties, such as mass. The amount of ejecta and explodibility at core-collapse is an inherent 3-dimensional problem \citep[e.g.][]{Muller+16}, which is mapped to 1-dimensional stellar properties. For this work, the BBH masses that can be formed  during collapse are critical for the formation of BBH mergers through SMT.

For the BH masses, the \citet{Fryer+12} \texttt{rapid} remnant mass prescription results in a mass gap in BH masses between $3\Msun$ and $5\Msun$, preventing their formation. For this reason, we use the \citet{Fryer+12} \texttt{delayed} prescription, since it allows for the formation of BHs in this mass range. By using this prescription, we know that if BHs in the $3{-}5\Msun$ range do not merge, it is caused by the stellar and binary physics and not by the remnant mass prescription, as shown in Section \ref{sec:BBH_merger_formation}.
Figure \ref{fig:other_SN_prescriptions} shows the supernova remnant mass prescriptions from \citet{Patton+20} and \citet{Sukhbold+16} with the \texttt{N20} and \texttt{W20} engines, respectively. Both remnant mass prescriptions more efficiently produce BBH mergers compared to the \citet{Fryer+12} \texttt{delayed} prescription in Figure \ref{fig:merger_pop}, shifting the main peak from $\Mmax \sim 15\Msun$ to $\Mmax \sim 10\Msun$. Above $\MBH \sim 15\Msun$, all BHs are formed through direct fallback in all remnant mass prescriptions in this work; thus, there are no noticeable differences between supernova prescriptions. Furthermore, the mass ratio distributions still peak at $\Mmin/\Mmax\sim1$, as both the primary and secondary are changed similarly by the remnant mass prescription.

We note that the \citet{Sukhbold+16} prescription, in the bottom row of Figure \ref{fig:other_SN_prescriptions}, leads to the formation of 3 BBH merger systems at $Z=0.45\Zsun$ for a sample size of 1.000.000 initial binaries. These are an artifact of the initial-final interpolation done across the \posydon grid, as the mass of the second-born BH increases compared to its pre-supernova mass increases slightly, and thus results in a system able to merge within the Hubble time. Given their low weight and occurrence, these do not drastically affect the overall SMT population properties.

In summary, the remnant mass prescription affects the low mass ($\MBH\simlt15\Msun$) regime of the BBH merger mass distributions, specifically altering the efficiency and the location of the low mass peak, but it does not drastically alter the allowed mass range at which BBH mergers are produced. Furthermore, a mass ratio of unity is preferred independent of remnant mass prescription for the populations without a natal kick.

\subsection{Natal kick populations}

\begin{figure*}
    \centering
    \includegraphics[width=\linewidth]{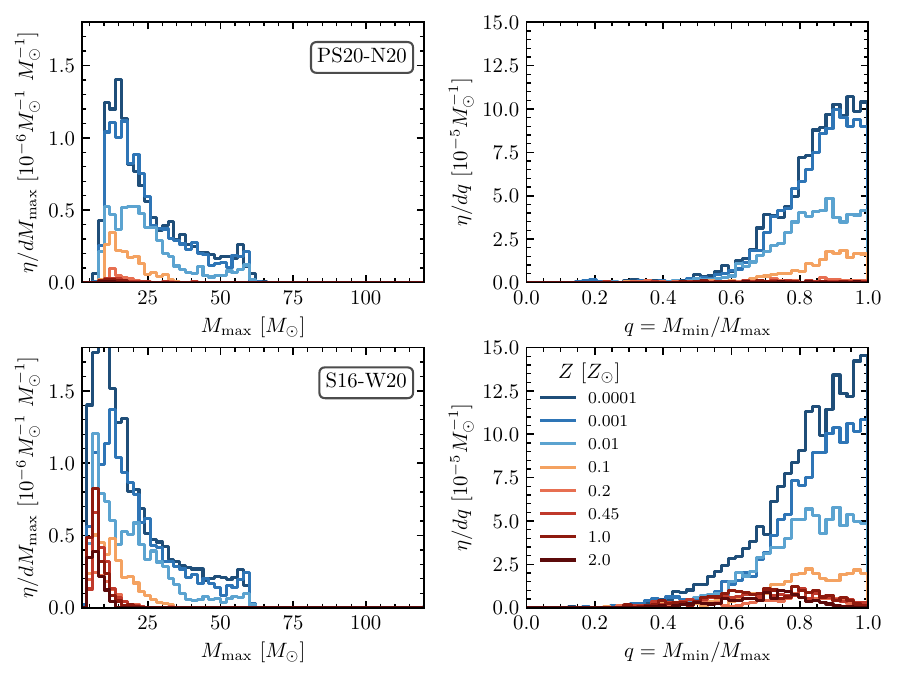}
    \caption{Mass and mass ratio distributions per metallicity for the \citet{Patton+20} remnant mass prescription with the \texttt{N20} engine \citep{Saio+88, Nomoto+88, Shigeyama+90} and the \citet{Sukhbold+16} prescription with the \texttt{W20} engine \citep{Woosley+97} on the top and bottom rows, respectively. Both prescriptions produce more BBHs compared to the \citet{Fryer+12} \texttt{delayed} prescription in Figure \ref{fig:merger_pop}. The colors indicate different metallicities, and a mass-scaled natal kick is applied.}
    \label{fig:other_SN_prescriptions_with_kick}
\end{figure*}

The population that receives a mass-scaled natal kick has a larger contribution of lower mass and more unequal mass ratio systems to the BBH merger rate, as discussed in Section \ref{sec:natal_kicks}. This sub-population is sensitive to the supernova remnant mass prescription, as shown in Figure \ref{fig:other_SN_prescriptions_with_kick}. The \texttt{PS20-N20} prescription has a lower contribution from high metallicities compared to \texttt{S16-W20} in Figure \ref{fig:other_SN_prescriptions_with_kick} and \citet{Fryer+12} \texttt{delayed} in Figure \ref{fig:merger_obs_kick}, most likely due to the formation of more neutron stars. The \texttt{S16-W20} prescription, on the other hand, produces many more low-mass BBH mergers and shifts the mass ratio distribution of the high-metallicity populations from $\Mmin/\Mmax\sim0.4$ to $0.7{-}0.8$. This indicates that the low mass peak in mass distribution and any peaks in the mass ratio distribution from kicked sub-populations are sensitive to the choice of the remnant mass prescription.

\end{appendix}
\end{document}